\documentclass[aps,prd,superscriptaddress,twocolumn]{revtex4}

\usepackage{amsfonts}
\usepackage{amsmath}
\usepackage{graphicx}
\usepackage{subfig}
\usepackage{color}
\usepackage{amssymb}
\usepackage[normalem]{ulem}
\usepackage{natbib}
\usepackage{soul}

\usepackage{float}

\usepackage[font=small, justification=centerlast, format=plain]{caption}

\begin{document}

\title{Electromagnetic fields induced by an electric charge near a Weyl semimetal}

\author{A. Mart\'{i}n-Ruiz}
\email{alberto.martin@nucleares.unam.mx}
\affiliation{Instituto de Ciencia de Materiales de Madrid, CSIC, Cantoblanco, 28049 Madrid, Spain.}

\affiliation{Instituto de Ciencias Nucleares, Universidad Nacional Aut\'{o}noma de M\'{e}xico, 04510 M\'{e}xico, Distrito Federal, M\'{e}xico}

\affiliation{Centro de Ciencias de la Complejidad, Universidad Nacional Aut\'{o}noma de M\'{e}xico, 04510 Ciudad de M\'{e}xico, M\'{e}xico}

\author{M. Cambiaso}
\email{mcambiaso@unab.cl}
\affiliation{Universidad Andres Bello, Departamento de Ciencias Fisicas, Facultad de Ciencias Exactas, Avenida Republica 220, Santiago, Chile}

\author{L. F. Urrutia}
\email{urrutia@nucleares.unam.mx}
\affiliation{Instituto de Ciencias Nucleares, Universidad Nacional Aut\'{o}noma de M\'{e}xico, 04510 M\'{e}xico, Distrito Federal, M\'{e}xico}

\begin{abstract}
Weyl semimetals (WSM) are a new class of topological materials that exhibit a bulk Hall effect and a chiral magnetic effect. The topological contribution of these unusual electromagnetic responses can be characterized by an axion term $\theta \textbf{E} \cdot \textbf{B}$ with space and time dependent axion angle $\theta (\textbf{r} ,t)$. In this paper we compute the electromagnetic fields produced by an electric charge near a topological Weyl semimetal with two Weyl nodes, in the equilibrium state, at zero electric chemical potential, and with broken time-reversal symmetry. We find that, as in ordinary metals and dielectrics, outside the WSM the electric field is mainly determined by the optical properties of the material. The magnetic field is, on the contrary, of topological origin due to the magnetoelectric effect of topological phases. We show that the magnetic field exhibits an interesting behavior above the WSM as compared with that induced above a topological insulator: the field lines begin at the surface and then 
end at the surface (but not at the same point). This distinctive  behavior of the magnetic field is an experimentally observable signature of the anomalous Hall effect in the bulk of the WSM. We discuss 
two experimental setups for testing our predictions of the induced magnetic field.
\end{abstract}

\maketitle

\section{Introduction}

Materials characterized by topological order, or simply topological materials have attracted great attention recently both from the theoretical and experimental fronts. The best studied of these are the topological insulators (TIs), which are characterized by a gapped bulk and protected boundary modes that are robust against  disorder \cite{Qi-Review, Hassan-Review}. Up to recent times, one usually associated topologically nontrivial properties with gapped systems; however, we have learned that gapless (semi)metallic states may be topologically nontrivial in  the same sense as gapped insulators. A particularly interesting state of matter is the topological Weyl semimetal (WSM), which may be thought of as a 3D analog of graphene. These are states characterized by phases with broken time-reversal (TR) or inversion (I) symmetry, whose electronic structure contains  a pair of Weyl nodes (band crossing points) in the Brillouin zone (BZ) provided the Fermi level is close to the Weyl nodes. WSMs possess protected gapless surface states on disconnected Fermi arcs with end points at the projection of the bulk nodes onto the surface BZ \cite{Armitage-Review}. The WSM phase was first theoretically predicted in pyrochlore iridates (such as Y$_2$Ir$_2$O$_7$) in 2011 \cite{TaAsTheo} and experimentally discovered in TaAs four years later \cite{TaAs-Huang, TaAs-Lv, TaAs-Xu, TaAs-Yang, TaAs-Xu2}.

Besides  their spectroscopic distinguishing features, topological phases also exhibit unusual electromagnetic (EM) responses that are  a direct macroscopic manifestation of the nontrivial topology of their band structure. It has been shown that the EM response of both topological insulators \citep{Qi-TFT, Essin-TFT, Wu-TFT} and Weyl semimetals \cite{Burkov-TFT, Burkov2-TFT, Tewari-TFT} is described by the so-called $\theta$ term in the EM action, $S _{\theta} \propto \int \theta (\textbf{r} , t) \, \textbf{E} \cdot \textbf{B} \, d ^{3} \textbf{r} dt$. For TIs, the only nonzero value compatible with TR symmetry is $\theta = \pi$, and thus has no effect on Maxwell equations in the bulk. Its only real effect, a half-quantized Hall effect on the sample's surfaces, becomes manifest only in the presence of surface magnetization. When TR and I symmetries are broken in the bulk, such as in a topological Weyl semimetal, the axion field $\theta$ may also acquire linear space and time dependence $\theta \left( \textbf{r} , t \right) = 2 \textbf{b} \cdot \textbf{r} - 2 b _{0} t$, where $2 \textbf{b}$ is the separation between the Weyl nodes in momentum space and $2 \hbar b _{0}$ is their energy offset.
Unlike the $\theta$ term for TIs, the analogous term for WSMs  modifies Maxwell equations in the bulk and thus has observable physical consequences, namely the anomalous Hall effect (AHE) and the chiral magnetic effect (CME). A number of physical effects, mainly optical, have been predicted on the basis of this theory. For example, the magneto-optical Faraday and Kerr rotation \cite{Kerr-Faraday/WSM} and the Casimir effect \cite{Casimir/WSM}, and the appearance of plasmon polaritons \cite{Plasmons/WSM} and helicons \cite{Helicons/WSM} at the sample's surface. In this paper we are concerned with a particular physical effect associated with the anomalous Hall effect.

One striking consequence of the $\theta$ term in topological insulators is the image magnetic monopole effect, namely, the appearance of a magnetic field that resembles the one produced by a magnetic monopole when an electric charge is put near the material's surface \cite{Qi-Monopole, Karch, MCU-GreenTI}. Physically, the monopole magnetic field is induced by a circulating Hall current on the TI surface, centered at the position of the charge projected onto the TI. In this paper we tackle the analogous effect in topological Weyl semimetals. To be precise, we investigate the electromagnetic fields induced by an electric charge above a WSM in the equilibrium state, at zero electric chemical potential, and with broken TR symmetry. We assume the charge to be located along the axis defined by the separation between the Weyl nodes in the BZ, i.e. near the surface without Fermi arcs. What is relevant in our configuration is that due to the magnetoelectric effect in WSMs, a magnetic field is induced. Outside the material, the magnetic field is noteworthy as it is not radial (as that produced by a magnetic monopole). Indeed, its physical origin is the anomalous Hall effect in the bulk, which as we will see, can be interpreted in terms of a family of $(2+1)$-dimensional subsystems parametrized by the coordinate along the nodal separation. Each subsystem exhibits a quantum-like Hall effect, such that a WSM can be effectively understood as a chain of 2D Dirac surface states.

The rest of this paper is organized as follows. In Sec. \ref{EM-Response} we briefly review the electromagnetic response of topological Weyl semimetals. The central part of this paper is presented in Sec. \ref{CalEMfields}, where we compute the EM fields produced by an electric charge above a WSM. In Sec. \ref{Force} we compute the interaction energy and the force that the material exerts upon the static charge. We close with a brief summary of our results and conclusions in Sec. \ref{Conclusions}, where we also discuss two possible experimental setups to eventually measure the resulting magnetic field. Appendix \ref{DetSol} contains the details of the calculation of the required scalar and vector potentials determining the electromagnetic fields. Throughout the paper we use Gaussian units.

\section{Electromagnetic response of Weyl semimetals} \label{EM-Response}

The low energy physics of a Weyl semimetal with two nodes is described by the linearized Hamiltonian \cite{Armitage-Review}
\begin{align}
H = v _{F} \hbar \tau ^{z} \boldsymbol{\sigma} \cdot \left( \textbf{k} + \tau ^{z} \textbf{b} \right) + \hbar \tau ^{z} b _{0} , \label{Hamiltonian}
\end{align}
where $v _{F}$ is the Fermi velocity and $\textbf{k} = - i \nabla$. The operator $\boldsymbol{\tau}$ describes the node degree of freedom, while $\boldsymbol{\sigma}$ describes the conduction-valence band degree of freedom. The separation of Weyl nodes in the BZ is governed by the 
broken symmetries in the bulk. A broken TR symmetry implies $\mathbf{b} \neq 0$ and this will produce a separation of the Weyl nodes in momentum by an amount $2 \mathbf{b}$, each node located at $\pm \mathbf{b}$. On the other hand, a broken I symmetry implies 
$b_0 \neq 0$ which will produce a separation of the Weyl nodes in energy, by an amount $2 \hbar b_0$.
The terms proportional to $b _{0}$ and $\textbf{b}$ in the Hamiltonian (\ref{Hamiltonian}) can be gauged away and it reduces to $H = v _{F} \hbar \tau ^{z} \boldsymbol{\sigma} \cdot \textbf{k}$. The chiral transformation in euclidean space $\psi ^{\prime} \to e^{-i \tau^z \theta/2} \psi$, with $\theta (\textbf{r},  t = i \tau) = 2 \textbf b \cdot \textbf r - 2 i b_0 \tau$ (and corresponding for $\psi^\dag$) indeed gauges away the terms $b_0 \tau^z$ and $\tau^z \boldsymbol{b} \cdot \boldsymbol{\sigma}$ but it also changes the integration measure in the path integral and thus the seeming chiral symmetry of the fermionic field  is broken, which is nothing more than the chiral anomaly. This gives rise to an unusual EM response described by an additional $\theta$ term in the action of the electromagnetic field \cite{Burkov-TFT, Burkov2-TFT, Tewari-TFT}
\begin{align}
S _{\theta} = \frac{\alpha}{4 \pi ^{2}} \int \theta (\textbf{r} , t) \, \textbf{E} \cdot \textbf{B} \, dt \, d ^{3} \textbf{r} , \label{ThetaTerm}
\end{align}
where $\alpha = e ^{2} / \hbar c$ is the fine-structure constant and $\theta (\textbf{r} , t) = 2 \textbf{b} \cdot \textbf{r} - 2 b _{0} t$ is the so-called axion field \cite{PQ, Wilczek}. Topological response of WSMs is thus described by an 
action similar to that of axion-electrodynamics.
 It is useful to compare this with the $\theta$ term in the effective action  of 3D topological insulators. In that case $\theta = \pi$ is the only nonzero value consistent with TR symmetry \citep{Qi-TFT, Essin-TFT, Wu-TFT}. The EM response of 3D TIs is rather simple, since the only nontrivial physical effect is to generate a half-quantized quantum Hall effect on the sample's surfaces. Indeed, a general method to describe the topological magnetoelectric effect in 3D TIs has been elaborated in Refs. \cite{MCU-GreenTI, Martin, MCU1, MCU2, MCU4} by means of Green's function techniques.

Unlike the $\theta$ term in 3D TIs, Eq. (\ref{ThetaTerm}) does modify Maxwell equations in the bulk of a Weyl semimetal and thus provides additional observable consequences. The physical manifestation of the Chern-Simons-like term (\ref{ThetaTerm}) can be best understood from the associated equations of motion. Varying the full action $S _{0} + S _{\theta}$, where $S _{0} = \frac{1}{8 \pi} \int \left[ \epsilon \textbf{E} ^{2} - (1 / \mu) \textbf{B} ^{2} \right] dt \, d ^{3} \textbf{r}$ is the usual  nontopological Maxwell action for electromagnetic fields in matter, we find that the axionic term (\ref{ThetaTerm}) changes two of the four Maxwell equations, i.e.
\begin{align}
\nabla \cdot \textbf{D} &= 4 \pi \left( \rho - \frac{\alpha}{2 \pi ^{2}} \textbf{b} \cdot \textbf{B} \right) , \label{Gauss}
\end{align}
and
\begin{align}
\nabla \times \textbf{H} - \frac{1}{c} \frac{\partial \textbf{D}}{\partial t} &= \frac{4 \pi}{c} \left( \textbf{J} + \frac{\alpha}{2 \pi ^{2}} c \textbf{b} \times \textbf{E} -  \frac{\alpha}{2 \pi ^{2}} b _{0} \textbf{B} \right) , \label{Ampere}
\end{align}
with the constitutive relations $\textbf{D} = \tilde{\epsilon} \, \textbf{E}$ and $\textbf{H} = \textbf{B} / \tilde{\mu}$. Faraday's law, $\nabla \times \textbf{E} = - c ^{-1} \partial \textbf{B} / \partial t$, and the equation stating the absence of magnetic monopoles, $\nabla \cdot \textbf{B} = 0$, remain unaltered. Here, $\tilde{\epsilon} = \epsilon + i \sigma _{xx} (\omega) / \omega$ and $\tilde{\mu} = 1 + \chi _{m}$, where $\epsilon $ is the static permittivity, $\sigma _{xx} (\omega)$ is the longitudinal conductivity and $\chi _{m}$ is the magnetic susceptibility that we  assume is negligible for  the WSM . 

In general, the electric current $\textbf{J}$ depends on both the electric and magnetic fields. As in ordinary metals, in the linear response regime, the electric field-dependent current is given by $\textbf{J} ^{\mbox{\scriptsize (E)}} = \sigma _{ij} (\omega) E _{j} \hat{\textbf{e}} _{i}$, where the frequency-dependent conductivity tensor $\sigma _{ij} (\omega)$ can be derived by using, for example, the semiclassical Boltzmann transport theory. In addition, if we have chiral fermions in a magnetic field with chemical potentials $\mu _{\mbox{\scriptsize L}}$ and $\mu _{\mbox{\scriptsize R}}$ for left- and right-handed fermions, respectively, there are two additional $\textbf{B}$-dependent current terms, namely, 
\begin{align}
\textbf{J} ^{\mbox{\scriptsize (B)}} = \frac{\alpha}{2 \pi ^{2}} \mu _{5} \textbf{B}  \quad , \quad \textbf{J}  _{5}  ^{\mbox{\scriptsize (B)}} = \frac{\alpha}{2 \pi ^{2}} \mu \textbf{B} , \label{JB}
\end{align}
where $\mu _{5} = ( \mu _{\mbox{\scriptsize L}} - \mu _{\mbox{\scriptsize R}}  ) / 2$ and $ \mu = ( \mu _{\mbox{\scriptsize L}} + \mu _{\mbox{\scriptsize R}} ) / 2$ are the chiral and the electric chemical potentials, respectively.

The most salient features of Weyl physics are fully contained in the inhomogeneous Maxwell equations (\ref{Gauss}) and (\ref{Ampere}). For example, the $\textbf{b}$-dependent terms encode the anomalous Hall effect that is expected to occur in a Weyl semimetal with broken TR symmetry \cite{AHE-Yang, AHE-Burkov, AHE-Grushin, AHE-Gorbar}. The $b _{0}$-dependent term that arises in Weyl semimetals with broken I symmetry, describes only one part of the celebrated chiral magnetic effect, namely, the generation of an electric current driven by an applied magnetic field. The second part of the CME is given by $\textbf{J} ^{\mbox{\scriptsize (B)}} $ in Eq. (\ref{JB}), which arises from an imbalance between chemical potentials of right-
and left-handed fermions. The total contribution to the CME current is:
\begin{equation}
\label{eq: cme curr}
\mathbf{J}_{\mathrm{CME}} =  -\frac{\alpha}{2 \pi ^{2}} \left( b _{0}  - \mu _{5} \right) \mathbf{B},
\end{equation}
that vanishes for  $b _{0}  = \mu _{5} $ in which case the WSM is said to be at the equilibrium state. On the other hand, $\mathbf{J}_5^{B}$ in Eq. (\ref{JB}) that is identified with the chiral separation effect, vanishes for $\mu = 0$,   condition that defines the neutrality point. For a detailed discussion of the chiral magnetic effect and the chiral separation effect see Ref. \cite{Landsteiner}. The vanishing of the CME in solid state context is addressed in Ref. \cite{CME-Vazifeh, CME-Kharzeev, CME-Ma}.

\section{Calculation of the EM fields} \label{CalEMfields}

\subsection{Statement of the problem}

An electric charge near the surface of a 3D TI induces a vortex Hall current (because of the in-plane component of the electric field produced by the charge)  generating a magnetic field that resembles the one produced by a magnetic monopole \cite{Qi-Monopole, Karch, MCU-GreenTI}. Similar image monopole is predicted when a charge is near the surface of linear magnetoelectric material \cite{Dzyaloshinskii}. In this paper we consider an electric charge near the surface of a topological Weyl semimetal.
Due to the  broken symmetries in the bulk,  additional nontrivial topological effects may result as compared to the case of the TIs. Specifically, we are concerned with the anomalous Hall effect of WSMs in the equilibrium state and at the neutrality point. Charge neutrality  can be attained for some WSMs under specific circumstances and it is not an unrealistic assumption. Theoretical and experimental studies involving WSMs at neutrality have been of considerable interest, as shown in the following cases. In Ref. \cite{Sbierski_2014} numerical calculations for transport properties are performed. Longitudinal and transversal conductivities and also topological Kerr and Faraday rotations were reported in Ref. \cite{Kerr-Faraday/WSM}. In Ref. \cite{Sun_2016} the authors report that TaAs exhibits strong spin Hall effect precisely at neutrality. In \cite{Xu_2016}, experimental confirmation of an optical conductivity with linear dependence in frequency (indicative of the Fermi level intersecting the Weyl nodes) for TaAs at $T=5 K$. A theoretical study of light propagation in a WSM finding unconventional electromagnetic modes was found \cite{Ferreiros_2016}. In Ref. \cite{Holder_2017} neutral WSMs in presence of strong disorder were studied finding that the residual conductivity is qualitatively larger than previously estimated. Realistic studies of transversal magnetoresistance and Shubnikovde Hass oscillations in WSM both away and at the neutrality point were considered in Ref. \cite{Klier_2017}. Finally, in Ref. \cite{Zhang_2018} the authors carried a theoretical \textit{ab initio} study of Berry curvature dipole in WSM. These examples show us that neutrality is not only a simplifying assumption, but rather a relevant one to be considered.

\begin{figure}
\includegraphics[scale=1]{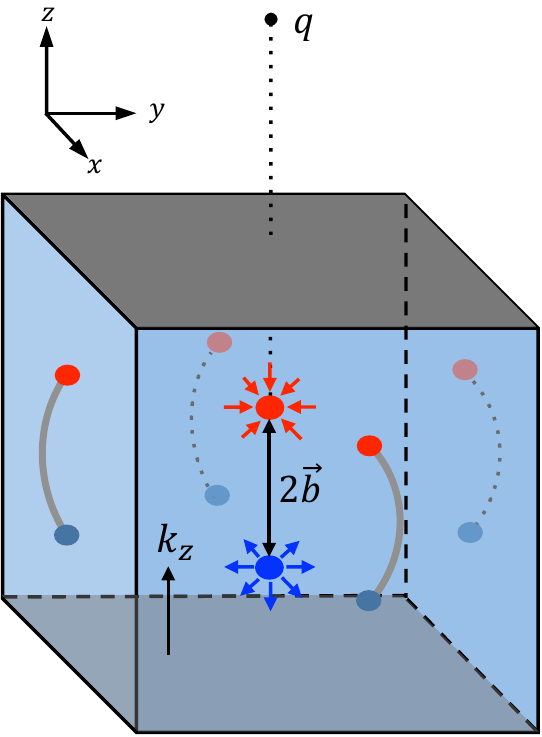}
\captionsetup{singlelinecheck=off}
\caption{Illustration of an electric charge above the surface of a Weyl semimetal. We also represent the $\textbf{k}$-space picture showing the location of the Weyl nodes (blue and red dots as sources and sinks of Berry curvature) along $k _{z}$ axis in the bulk BZ and the Fermi arcs (lines ending at the projection of the Weyl nodes) on the surface BZ.} \label{figure}
\end{figure}

Let us consider the geometry  shown in Fig. \ref{figure}. The lower half-space ($z<0$) is occupied by a topological Weyl semimetal with a pair of nodes separated along the $k _{z}$-direction in the bulk BZ, while the upper half-space ($z>0$) is occupied by a dielectric fluid. An electric charge is brought near  the surface that does not support Fermi-arc electronic states, in this case the $xy$-plane for $\textbf{b} = b \hat{\textbf{e}} _{z}$. Being this a static problem, it is appropriate to neglect all frequency dependence on the conductivities, such that the EM response of the WSM is fully captured by Eqs. (\ref{Gauss}) and (\ref{Ampere}), with $b _{0} = \mu _{5}$ and $\mu = 0$. Since $\theta (z = 0) = 0$, there are no surface currents, and the resulting material is just a bulk Hall medium with current responses given by the transverse Hall conductivity
\begin{align}
\sigma _{xy} = - \sigma _{yx} = \frac{e ^{2} b}{2 \pi ^{2} \hbar} . \label{sigmaxy}
\end{align}
The analogous problem of a charge located in front of a surface that supports Fermi arcs would also be of interest. However, from a practical point of view, we start from the assumption that the WSM phase has been properly characterized, such that the surfaces with/without Fermi arcs have been identified and then we can choose the configuration depicted in Fig. \ref{figure}. In fact, when a WSM phase is produced from a Dirac semimetal by applying an external magnetic field, the separation between nodes will be along the field direction and thus the identification of the faces supporting surface states is possible. Hereafter we concentrate on the surface without Fermi arcs, and we left the complementary problem for future investigations.

For the sake of generality, in section \ref{GenSol} we solve Maxwell equations (\ref{Gauss}) and (\ref{Ampere}) by considering 
two semi infinite bulk Hall materials, characterized by the parameters $(\epsilon _{1} , b _{1})$ for $z<0$ and $(\epsilon _{2} , b _{2})$ for $z>0$,  separated by the surface $z = 0$. The inhomogeneity in $\epsilon (\textbf{r})$ and $\sigma _{xy} (\textbf{r})$ is therefore limited to a finite discontinuity across the interface. Our results correctly reproduce the ones reported in Ref. \cite{ChiralMatter} for an infinite chiral medium, and the well-known electrostatic field produced by a charge near a dielectric medium \cite{Schwinger} as well. In the last section \ref{PartSol} we take the limit $b _{2} = 0$, which yields the electromagnetic fields produced by an electric charge in a dielectric fluid above the surface of a WSM.

\subsection{General solution and consistency checks} \label{GenSol}

Since the homogeneous Maxwell equations relating the potentials to the fields are not modified by the $\theta$ term, the static electric and magnetic fields can be written in terms of the scalar $\Phi$ and vector $\textbf{A}$ potentials according to $\textbf{E} = - \nabla \Phi$ and $\textbf{B} = \nabla \times \textbf{A}$. In the Coulomb gauge $\nabla \cdot \textbf{A} = 0$, for a pointlike electric charge of strength $q$ at $\textbf{r} ^{\prime} = z ^{\prime} \hat{\textbf{e}} _{z}$ with $z ^{\prime} >0$ (that is the charge lies in  medium 2), the electromagnetic potentials satisfy the equations 
\begin{align}
- \nabla \cdot \left[ \epsilon (z) \nabla \Phi \right] + \frac{4 \pi}{c} \sigma _{xy} (z) \, \hat{\textbf{e}} _{z} \cdot \nabla \times \textbf{A} &=  4 \pi \rho (\textbf{r}) , \label{Maxwell1} \\ - \nabla ^{2} \textbf{A} + \frac{4 \pi}{c} \sigma _{xy} (z) \, \hat{\textbf{e}} _{z} \times \nabla \Phi &= 0 \label{Maxwell2} ,
\end{align}
where $\rho (\textbf{r}) = q \delta (\textbf{r} - \textbf{r} ^{\prime})$ is the charge density. To obtain the general solution for the EM potentials, we must solve equations (\ref{Maxwell1}) and (\ref{Maxwell2}) in the bulk Hall systems and satisfy the appropriate boundary conditions. Working in cylindrical coordinates $(\rho, \varphi ,z)$ to exploit the axial symmetry of the problem, we introduce the reduced scalar potential $\phi (z , z ^{\prime} ; k _{\perp})$ through the $2+1$ representation
\begin{align}
\Phi (\textbf{r}) = 4 \pi \int \frac{ d ^{2} \textbf{k} _{\perp}}{(2 \pi) ^{2}} e ^{i \textbf{k} _{\perp} \cdot \boldsymbol{\rho}} \phi (z , z ^{\prime} ; k _{\perp}) , \label{RedEscPot}
\end{align}
where $\textbf{k} _{\perp} = (k _{x} , k _{y})$ and $\boldsymbol{\rho} = (x ,y)$ are the momentum and position parallel to the $xy$-plane. Expressing the area element in polar coordinates, $d ^{2} \textbf{k} _{\perp} = k _{\perp} dk _{\perp} d \varphi$, and choosing the $k _{x}$ axis in the direction of the vector $\boldsymbol{\rho}$, then $\textbf{k} _{\perp} \cdot \boldsymbol{\rho} = k _{\perp} \rho \cos \varphi$ and the angular integration can be performed to obtain
\begin{align}
\Phi (\rho , z) = 2 \int _{0} ^{\infty} k _{\perp} \, J _{0} (k _{\perp} \rho) \phi (z , z ^{\prime} ; k _{\perp}) dk _{\perp} , \label{RedEscPot2}
\end{align}
where $J _{n}$ is the $n$th order Bessel function of the first kind. Inserting this ansatz into the equations of motion and assuming the axial symmetry for the vector potential, we introduce the analogous $2+1$ dimensional representation
\begin{align}
\Psi (\rho , z) = 2 \int _{0} ^{\infty} k _{\perp} \, J _{1} (k  _{\perp} \rho) \psi (z , z ^{\prime} ; k _{\perp}) dk _{\perp} , \label{RedVecPot}
\end{align}
which defines the vector potential through $\textbf{A} = \Psi (\rho , z) \hat{\textbf{e}} _{\varphi}$, choice that naturally satisfies the Coulomb gauge $\nabla \cdot \textbf{A} = \rho ^{-1} \partial _{\varphi} \Psi = 0$. The problem now consists in determining the reduced functions $\phi$ and $\psi$. Inserting the above $2+1$ representations into Eqs. (\ref{Maxwell1}) and (\ref{Maxwell2}) we obtain
\begin{align}
- \frac{\partial}{\partial z} \left( \epsilon \frac{\partial \phi}{\partial z} \right) + k ^{2} _{\perp} \epsilon \phi + \frac{4 \pi}{c} k _{\perp} \sigma _{xy} \psi &= q \delta (z - z ^{\prime}) , \label{MaxRed1} \\ - \frac{\partial ^{2} \psi}{\partial z ^{2}} + k ^{2} _{\perp} \psi - \frac{4 \pi}{c} k _{\perp} \sigma _{xy} \phi &= 0 , \label{MaxRed2} 
\end{align}
where we have expressed the charge density as $\rho (\textbf{r}) = \frac{q}{2 \pi} \delta (z - z ^{\prime}) \int _{0} ^{\infty} k _{\perp} J _{0} (k _{\perp} \rho) dk _{\perp}$. Here, we omit the $z$-dependence of the dielectric function $\epsilon$ and the Hall conductivity $\sigma _{xy}$ for brevity. The differential equations (\ref{MaxRed1}) and (\ref{MaxRed2}), along with the appropriate boundary conditions at the interface $z = 0$ and at the singular point $z = z ^{\prime}$, constitute a complete boundary value problem. 

Before dealing with the solutions of the above equations let us comment on the  effects on the induced electric and magnetic fields  when we interchange the  position of the Weyl nodes in momentum space. In the following we momentarily call the solutions to Eqs. (\ref{MaxRed1}) and (\ref{MaxRed2}) as $\phi_\mathbf{b}$ and $\psi_\mathbf{b}$. There are two possibilities of arranging the Weyl nodes according to whether the source (sink) of the  Berry curvature is
located at  $+\mathbf{b} \,\, (-\mathbf{b})$ or the other way around, which amounts to changing $\mathbf{b} \to - \mathbf{b}$ in our equations. This interchange implies $\sigma _{xy} \to - \sigma _{xy}$ in Eqs. (\ref{MaxRed1}) and (\ref{MaxRed2}), with new solutions we now denote by $\phi_{-\mathbf{b}}$ and $\psi_{-\mathbf{b}}$. However, by the way in which these equations are coupled we obtain $\phi_{-\mathbf{b}} = \phi_\mathbf{b}$ and 
$\psi_{-\mathbf{b}} = - \psi_\mathbf{b}$. In other words, the electrostatic potential remains the same, but the vector potential (and thus the magnetic field) flips sign under the interchange of the Weyl nodes in momentum space.

To solve equations (\ref{MaxRed1}) and (\ref{MaxRed2}), we employ standard techniques of electromagnetism \cite{Schwinger}. Leaving the details of the  calculations for Appendix \ref{DetSol}, we obtain the following expressions for the reduced functions beneath the surface ($z<0$)
\begin{widetext}
\begin{align}
\phi _{z<0} = & \; \frac{q}{\epsilon _{1} Q} e ^{\alpha _{1} z - \alpha _{2} z ^{\prime}} \Big\{ \left( \epsilon _{1} \alpha _{1} + \epsilon _{2} \alpha _{2} \right) \cos \left( \beta _{1} z - \beta _{2} z ^{\prime} \right) + \left( \epsilon _{1} \beta _{1} + \epsilon _{2} \beta _{2} \right) \sin \left( \beta _{1} z - \beta _{2} z ^{\prime} \right)  + \left( \epsilon _{1} - \epsilon _{2} \right) \notag \\ & \times \cos \left( \beta _{1} z \right) \left[ \alpha _{2} \cos \left( \beta _{2} z ^{\prime} \right) - \beta _{2} \sin \left( \beta _{2} z ^{\prime} \right) \right] \Big\} , \label{RedEscz<0} \\ \psi _{z<0} = & \; \frac{q}{Q} e ^{\alpha _{1} z - \alpha _{2} z ^{\prime}} \Big\{ \left( \epsilon _{1} \beta _{1} + \epsilon _{2} \beta _{2} \right) \cos \left( \beta _{1} z - \beta _{2} z ^{\prime} \right) - \left( \epsilon _{1} \alpha _{1} + \epsilon _{2} \alpha _{2} \right) \sin \left( \beta _{1} z - \beta _{2} z ^{\prime} \right) + \left( \epsilon _{1} - \epsilon _{2} \right) \notag \\ & \times \sin \left( \beta _{1} z \right) \left[ \beta _{2} \sin \left( \beta _{2} z ^{\prime} \right) - \alpha _{2} \cos \left( \beta _{2} z ^{\prime} \right) \right] \Big\} , \label{RedVecz<0}
\end{align}
and, above the surface ($z>0$), we obtain
\begin{align}
\phi _{z>0} = & \; \frac{q}{2 \epsilon _{2} r _{2} ^{2}} e ^{- \alpha _{2} \vert z - z ^{\prime} \vert} \Big[ \alpha _{2} \cos \left( \beta _{2} \vert z - z ^{\prime} \vert \right) - \beta _{2} \sin \left( \beta _{2} \vert z - z ^{\prime} \vert \right) \Big] - \frac{q (\epsilon _{1} - \epsilon _{2})}{2 \epsilon _{2} Q} e ^{- \alpha _{2} (z + z ^{\prime})} \Big\{ \alpha _{1} \cos \left[ \beta _{2} \left( z - z ^{\prime} \right) \right] \notag \\ & + \beta _{1} \sin \left[ \beta _{2} \left( z - z ^{\prime} \right) \right] \Big\} + \frac{q}{2 \epsilon _{2} r _{2} ^{2} Q} e ^{- \alpha _{2} (z + z ^{\prime})} \Big\{ \Gamma \cos \left[ \beta _{2} \left( z + z ^{\prime} \right) \right] - \Delta \sin \left[ \beta _{2} \left( z + z ^{\prime} \right) \right] \Big\} , \label{RedEscz>0} \\ \psi _{z>0} = & \; \frac{q}{2 r _{2} ^{2}} e ^{- \alpha _{2} \vert z - z ^{\prime} \vert} \Big[ \beta _{2} \cos \left( \beta _{2} \vert z - z ^{\prime} \vert \right) + \alpha _{2} \sin \left( \beta _{2} \vert z - z ^{\prime} \vert \right) \Big] + \frac{q (\epsilon _{1} - \epsilon _{2})}{2 Q} e ^{- \alpha _{2} (z + z ^{\prime})} \Big\{ \beta _{1} \cos \left[ \beta _{2} \left( z - z ^{\prime} \right) \right] \notag \\ & - \alpha _{1} \sin \left[ \beta _{2} \left( z - z ^{\prime} \right) \right] \Big\} + \frac{q}{2 r _{2} ^{2} Q} e ^{- \alpha _{2} (z + z ^{\prime})} \Big\{ \Delta \cos \left[ \beta _{2} \left( z + z ^{\prime} \right) \right] + \Gamma \sin \left[ \beta _{2} \left( z + z ^{\prime} \right) \right] \Big\} , \label{RedVecz>0}
\end{align}
\end{widetext}

\begin{figure*}[t]
\subfloat[]{\includegraphics[scale = 0.23]{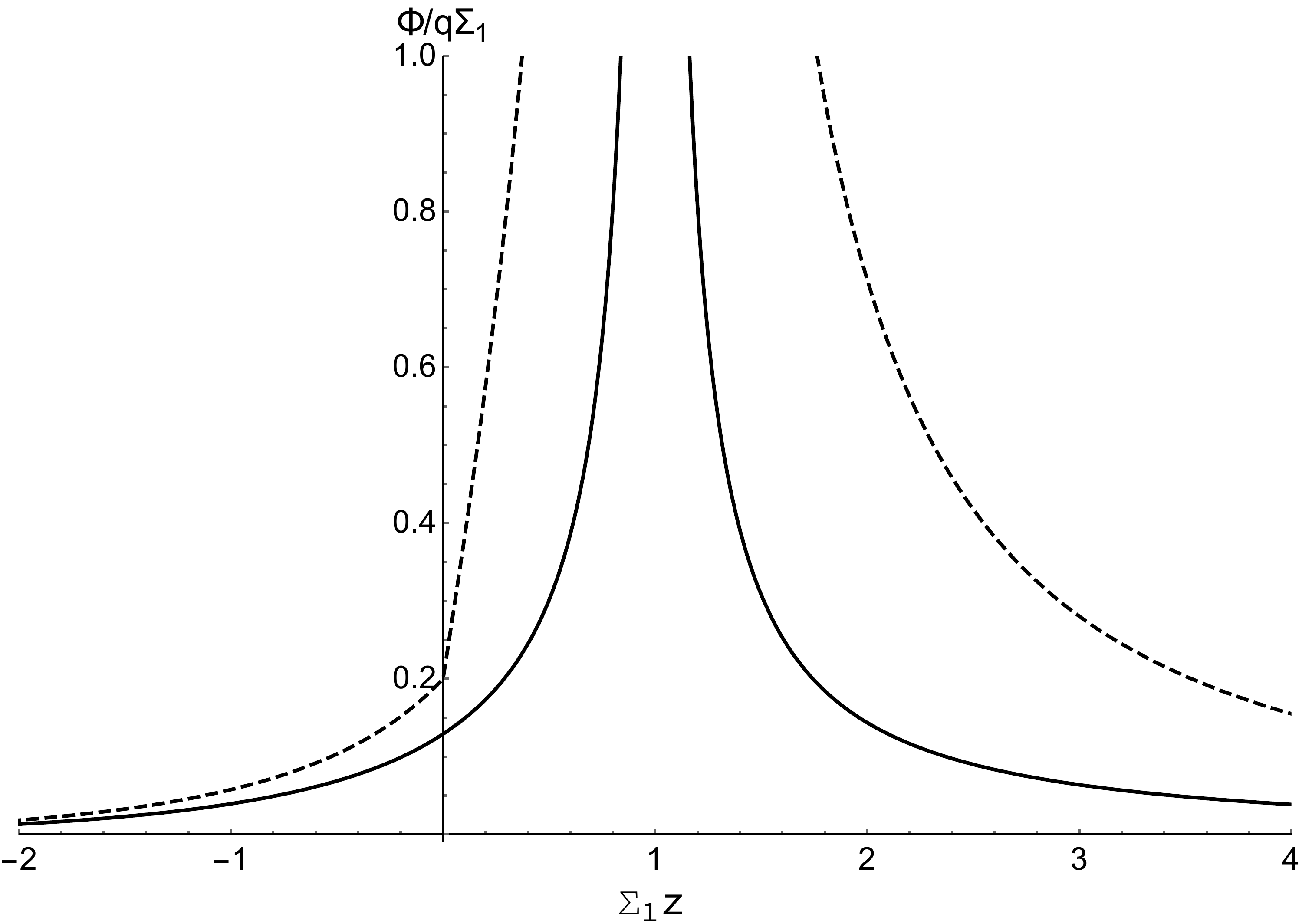}} \hfill
\subfloat[]{\includegraphics[width = 2in]{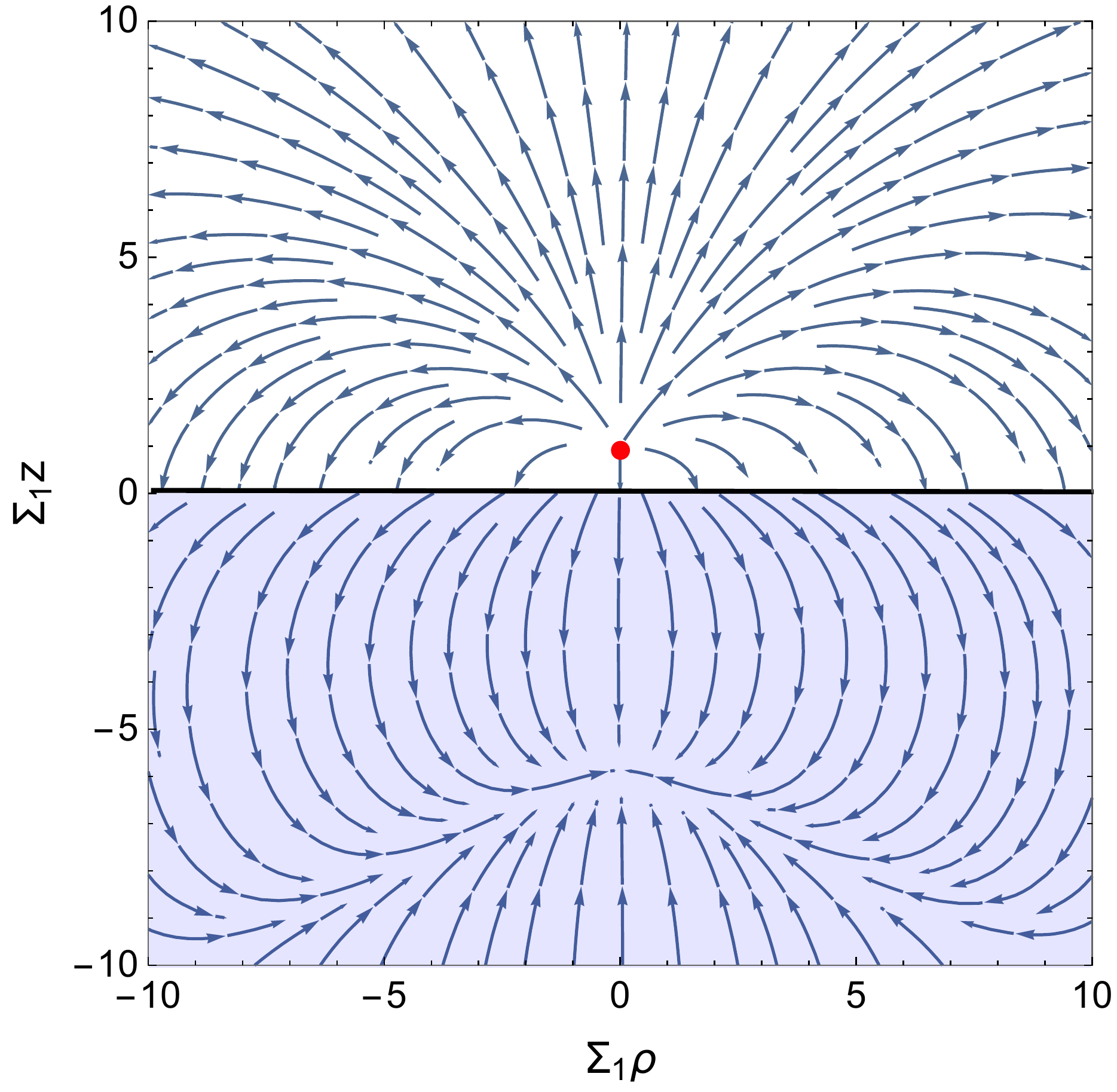}} \hfill
\subfloat[]{\includegraphics[width = 2in]{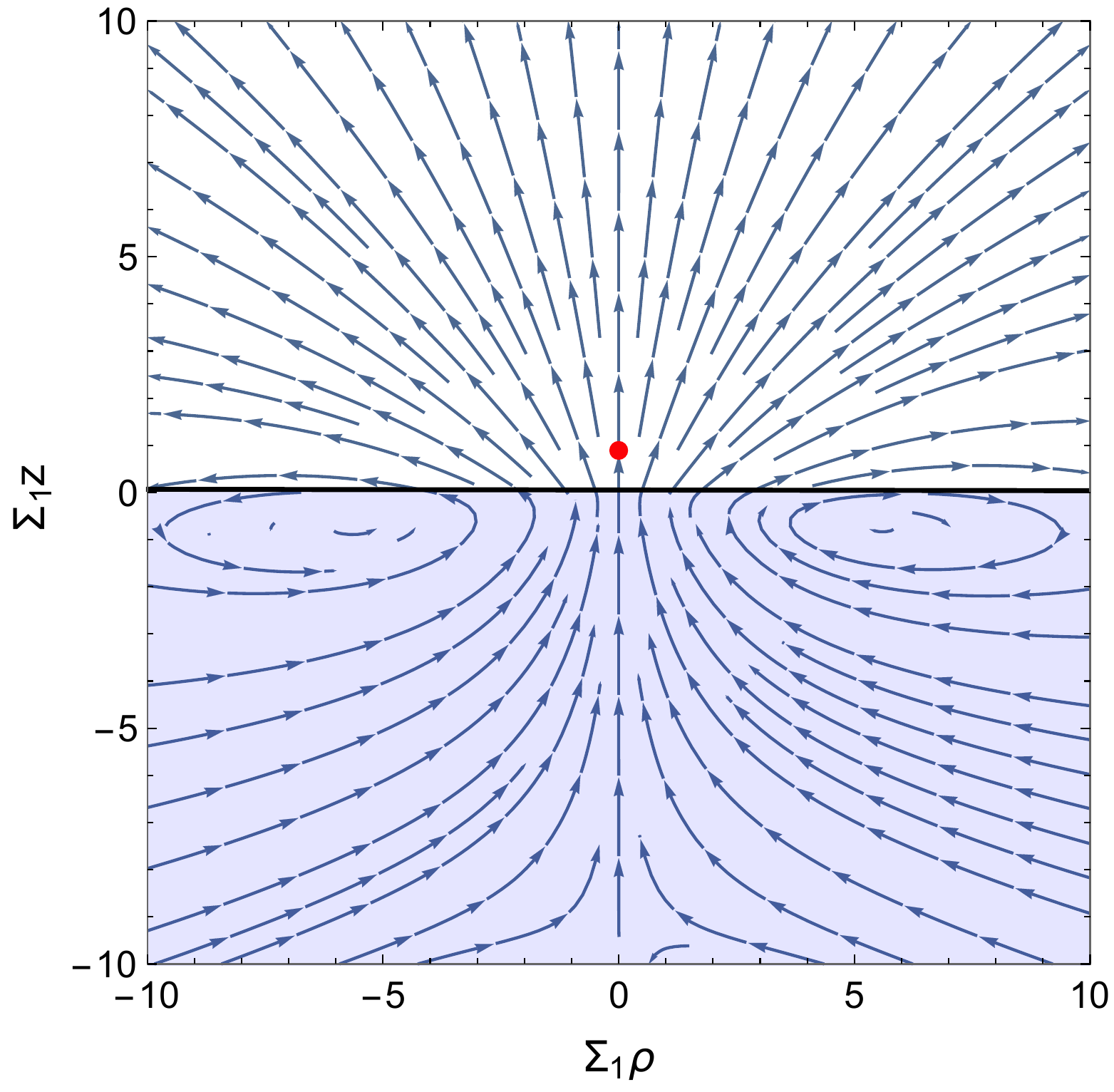}}
\caption{\small a) Plot of the electric potential $\Phi$ (in units of $q \Sigma _{1}$) as a function of the dimensionless distance $\Sigma _{1} z$ for $\rho = 0$, when the WSM is characterized by $\epsilon _{1} = 6$. The continuous line is for $\epsilon _{2} = 6$ and the dashed line for $\epsilon _{2} = 1$. b) and c) illustrates the electric and magnetic fields (in units of $q \Sigma _{1} ^{2}$), respectively, as a function of $\Sigma _{1} z$ and $\Sigma _{1} \rho$. The red sphere marks the position of the electric charge, at $z ^{\prime} = 1 / \Sigma _{1}$. The field lines in b) and c) correspond to the choice $b>0$. Under the interchange of the Weyl nodes in momentum space,  $\mathbf{b} \to -\mathbf{b}$, the field lines of the electric field remain invariant, while the field lines of the magnetic field are inverted.} \label{EMPlots}
\end{figure*}
where we have defined
\begin{align}
\Gamma & = \alpha _{2} \left( \epsilon _{2} r _{2} ^{2} - \epsilon _{1} r _{1} ^{2} \right) + \beta _{2} \left( \epsilon _{1} + \epsilon _{2} \right) \left( \alpha _{1} \beta _{2} - \beta _{1} \alpha _{2} \right) , \notag \\ \Delta & = \beta _{2} \left( \epsilon _{2} r _{2} ^{2} - \epsilon _{1} r _{1} ^{2} \right) - \alpha _{2} \left( \epsilon _{1} + \epsilon _{2} \right) \left( \alpha _{1} \beta _{2} - \beta _{1} \alpha _{2} \right) , \notag \\ Q &= \epsilon _{1} r _{1} ^{2} + \epsilon _{2} r _{2} ^{2} + \left( \epsilon _{1} + \epsilon _{2} \right) \left( \alpha _{1} \alpha _{2} + \beta _{1}  \beta _{2} \right) , \label{Defs}
\end{align}
and $r _{j} ^{2} = k _{jz} k _{j z} ^{\ast}$. Here, $k _{j z} = \alpha _{j} ( k _{\perp}) + i \beta _{j} ( k _{\perp})$ is the complex wave number in the medium $j$, with
\begin{align}
\alpha _{j} ( k _{\perp}) = \sqrt{\frac{k _{\perp}}{2} \left( \sqrt{k ^{2} _{\perp} + \Sigma _{j} ^{2}} + k _{\perp} \right)} , \notag \\ \beta _{j} ( k _{\perp}) = \sqrt{\frac{k _{\perp}}{2} \left( \sqrt{k ^{2} _{\perp} + \Sigma _{j} ^{2}} - k _{\perp} \right)} , \label{kappa}
\end{align}
and $\Sigma _{j} = \frac{4 \pi}{c} \frac{\sigma _{xy} ^{j}}{\epsilon _{j}}$ is an effective bulk Hall conductivity (with dimensions of inverse length). The imaginary part $\beta _{j}$ of $k _{jz}$ implies that the electromagnetic fields are attenuated in the bulk, as in ordinary metals.

The final expressions for the scalar and vector potentials in coordinate representation are obtained by inserting the reduced functions (\ref{RedEscz<0})-(\ref{RedVecz>0}) into the $2+1$ representations (\ref{RedEscPot2})-(\ref{RedVecPot}) and computing the $k _{\perp}$-integrals. Below we present two consistency checks of our results. 

First, we consider the limit in which the two materials are topologically trivial (i.e., with vanishing bulk Hall conductivities, $\Sigma _{1} = \Sigma _{2} = 0$), with however $\epsilon _{1} \neq \epsilon _{2}$. In this case we find that $k _{1z} = k _{2z} = k _{\perp}$, thus yielding $Q = 2 k ^{2} _{\perp} \left( \epsilon _{1} + \epsilon _{2} \right)$, $\Gamma = k ^{3} _{\perp} \left( \epsilon _{2} - \epsilon _{1} \right)$ and $\Delta = 0$. Therefore, the reduced scalar potential can be written 
as
\begin{align}
\phi &= \frac{q}{2 \epsilon _{2} k _{\perp}} \left[ e ^{- k _{\perp} \vert z - z ^{\prime} \vert} - \frac{\epsilon _{1} - \epsilon _{2}}{\epsilon _{1} + \epsilon _{2}} e ^{- k _{\perp} \vert z \vert} e ^{- k _{\perp} \vert z ^{\prime} \vert} \right] ,
\end{align}
which we recognize as that of an electric charge in front of a dielectric interface \cite{Schwinger}. Besides, 
this limit yields $\psi _{z>0} = \psi _{z<0} = 0$, as expected, since there is no magnetoelectric effect in the absence of the $\theta$ term.

Second, we consider the case in which the electric charge is embedded in an infinite chiral 
medium, namely $\epsilon_1 = \epsilon_2 \equiv \epsilon$ and $\Sigma_1 = \Sigma_2 \equiv \Sigma$, then  $Q = 4 \epsilon k _{\perp} \sqrt{k _{\perp} ^{2} + \Sigma ^{2}}$ and $\Gamma = \Delta = 0$. The  resulting scalar and vector potentials due to a charge $q$ located
at $z ^{\prime} = 0$ are:
\begin{align}
\Phi &= \frac{q}{\epsilon} \int _{0} ^{\infty} \!\!\! k_\perp J _{0} \left( k_\perp \rho \right) \left[ \frac{1}{2 k_z} e ^{- k_z \vert z \vert} + \frac{1}{2 k_z ^{\ast}} e ^{- k_z ^{\ast} \vert z \vert} \right] dk_\perp ,  \notag \\ \textbf{A} &= q \!\int _{0} ^{\infty}\!\!\! k_\perp J _{1} (k_\perp \rho) \left[ \frac{i}{2 k_z} e ^{- k_z \vert z \vert} - \frac{i}{2 k_z ^{\ast}} e ^{- k_z ^{\ast} \vert z \vert} \right] dk_\perp \, \hat{\textbf{e}} _{\varphi} ,\label{VectorApp}
\end{align}
where $k_z \equiv  \alpha + i \beta$ and $\alpha$, $\beta$ are given by Eq. (\ref{kappa}) with $\Sigma_j = \Sigma$.
As expected, these expressions coincide with the ones obtained from the Green's functions in Ref. \cite{ChiralMatter}.

\subsection{EM fields induced by a charge near a WSM} \label{PartSol}

The case of an electric charge located in a dielectric fluid, above the surface of a topological Weyl semimetal, as shown in Fig. \ref{figure}, is described by the reduced functions (\ref{RedEscz<0})-(\ref{RedVecz>0}) in the limit $\Sigma _{2} = 0$. First we discuss the resulting electric field. Taking $\beta _{2} = 0$ in Eqs. (\ref{RedEscz<0}) and (\ref{RedEscz>0}) and inserting the result into the $2+1$ representation (\ref{RedEscPot2}) we find that, in coordinate representation, the electrostatic potential beneath the surface becomes
\begin{align}
\Phi _{z<0} = & \; 2q \int _{0} ^{\infty} \frac{\left( \alpha _{1} + k _{\perp} \right) \cos \left( \beta _{1} z \right) + \beta _{1} \sin \left( \beta _{1} z \right)}{\epsilon _{1} \left( \alpha _{1} ^{2} + \beta _{1} ^{2} \right) + \epsilon _{2} k ^{2} _{\perp} + k _{\perp} \alpha _{1} \left( \epsilon _{1} + \epsilon _{2} \right)} \notag \\ & \times k _{\perp} J _{0} (k _{\perp} \rho) e ^{\alpha _{1} z - k _{\perp} z ^{\prime}} dk _{\perp} , \label{EscPotWSM}
\end{align}
and, above the surface, we find
\begin{align}
\Phi _{z>0} &= \frac{q}{\epsilon _{2}} \frac{1}{\sqrt{\rho ^{2} + (z - z ^{\prime}) ^{2}}} + \frac{q}{\epsilon _{2}} \frac{\epsilon _{2} - \epsilon _{1}}{\epsilon _{2} + \epsilon _{1}} \frac{1}{\sqrt{\rho ^{2} + (z + z ^{\prime}) ^{2}}} \notag \\ & \phantom{=} - \frac{2 q \epsilon _{1}}{\epsilon _{1} + \epsilon _{2}}  \int _{0} ^{\infty} \!\!\!\! \frac{\alpha _{1} ^{2} + \beta _{1} ^{2} - k ^{2} _{\perp}}{\epsilon _{1} \! \left( \alpha _{1} ^{2} + \beta _{1} ^{2} \right) + \epsilon _{2} k ^{2} _{\perp} +  k _{\perp} \alpha _{1} \! \left( \epsilon _{1} \! + \! \epsilon _{2} \right)} \notag \\ & \phantom{=} \times J _{0} (k _{\perp} \rho) e ^{- k _{\perp} ( z + z ^{\prime} )} dk _{\perp} . \label{EscPotDielectric}
\end{align}
We observe that in the dielectric fluid ($z>0$), the electric potential can be interpreted as due to the original electric charge of strength $q$ at $z ^{\prime}$, an image electric charge of strength $q (\epsilon _{2} - \epsilon _{1}) / (\epsilon _{2} + \epsilon _{1})$ at $-z ^{\prime}$, and an additional term arising from the nontrivial topology of the WSM. Inside the Weyl semimetal ($z<0$), the electric potential has no  simple interpretation. In Fig. \ref{EMPlots}a we plot the electrostatic potential $\Phi$ (in units of $q \Sigma _{1}$) as a function of the dimensionless distance $\Sigma _{1} z$, for $\rho = 0$ and $z ^{\prime} = 1 / \Sigma _{1}$. 
Consider the reference value $\epsilon _{1} \sim 6 $ appropriate for dielectric constant of the Weyl semimetal TaAs  \cite{TaAs-Huang, TaAs-Lv, TaAs-Xu, TaAs-Yang, TaAs-Xu2}. In Fig. \ref{EMPlots}a the continuous line represents the case when the semispace above the WSM is also filled with a dielectric medium of $\epsilon _{2} = 6$, and the dashed line represents the case when the space above the WSM is vacuum, namely $\epsilon _{2} = 1$. As expected, we observe that the electrostatic potential is attenuated inside the WSM due to the metallic character of the material. 
\begin{figure*}[t]
\centering
\includegraphics[scale=0.5]{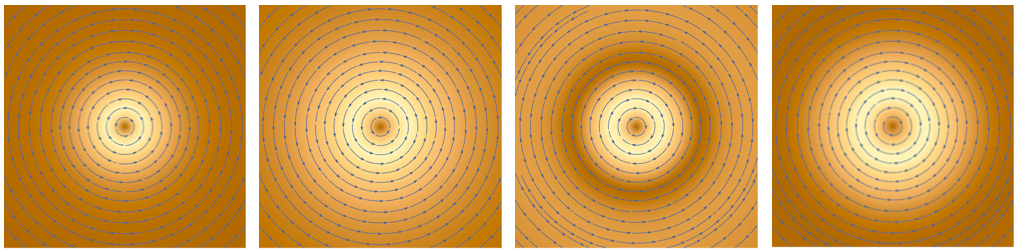}
\caption{\small Stream density plots of the bulk Hall current $\textbf{J} _{\mbox{\scriptsize bHall}}$ (in units of $q \sigma _{xy} \Sigma _{1} ^{2}$) for $z ^{\prime} = 1 / \Sigma _{1}$, $\epsilon _{1} = 6$, $\epsilon _{1} =1$, and $\Sigma _{1} z = -1, -2, -3.5 , -4$ (from left to right).}
\label{HallCurrentPlots}
\end{figure*}
From Eq. (\ref{EscPotWSM}), one can further see that, in the limit $\epsilon _{1} \rightarrow \infty$, we obtain that $\Phi _{z<0} = 0$, as in a perfect conductor. Even in this simplified model of a WSM this decay of the electrostatic potential inside the material reflects an additional contribution to the screening length of the material \cite{ZangWill}, which is not the penetration depth as defined for electromagnetic waves. A proper 
estimation of these relevant parameters demands to consider a more realistic model for the WSM and deserves further investigation.

The electric field is obtained  from the electrostatic potential (\ref{EscPotWSM})-(\ref{EscPotDielectric}) as $\textbf{E} = - \nabla \Phi$. In Fig. \ref{EMPlots}b we illustrate the electric field $\textbf{E}$ (in units of $q \Sigma _{1} ^{2}$) generated by an electric charge in vacuum ($\epsilon _{2} = 1$) at $z ^{\prime} = 1 / \Sigma _{1}$ (red sphere) close to the WSM TaAs as a function of the dimensionless coordinates $\Sigma _{1} \rho$ and $\Sigma _{1} z$. We observe that the electric field outside the WSM is similar to that generated by the original electric charge, with deviations close to the interface due to the screening of the field inside the material. In fact, this electric field is practically indistinguishable from that produced by an electric charge close to an ordinary metal or a dielectric. Nevertheless, the electric field beneath the surface is more complicated than in the nontopological cases. For example, the electric field within a uniform and isotropic dielectric is a radially directed field (with the charge outside the material as its source); while the field inside an ordinary metal is zero. In the present case, as shown in Fig. \ref{EMPlots}b, the electric field is remarkably different 
as evidenced by the curved field lines inside.

Now we discuss the induced magnetic field. The vector potential is given by $\textbf{A} = \Psi \hat{\textbf{e}} _{\varphi}$, with the function $\Psi$ defined by Eq. (\ref{RedVecPot}) and the reduced function $\psi$ given by Eqs. (\ref{RedVecz<0}) and (\ref{RedVecz>0}) in the limit $\beta _{2} = 0$. In coordinate representation, the function $\Psi$ beneath the surface is
\begin{align}
\Psi _{z<0} = & \, 2 q \epsilon _{1} \int _{0} ^{\infty}  \frac{ \beta _{1} \cos \left( \beta _{1} z \right) - \left( \alpha _{1} + k _{\perp} \right) \sin \left( \beta _{1} z \right) }{\epsilon _{1} \left( \alpha _{1} ^{2} + \beta _{1} ^{2} \right) + \epsilon _{2} k ^{2} _{\perp} + k _{\perp} \alpha _{1} \left( \epsilon _{1} + \epsilon _{2} \right)} \notag \\ & \times k _{\perp} J _{1} (k _{\perp} \rho) e ^{\alpha _{1} z - k _{\perp} z ^{\prime}} dk _{\perp} ,
\end{align}
and, above the surface, we obtain
\begin{align}
\Psi _{z>0} = & \, 2 q \epsilon _{1} \int _{0} ^{\infty} \frac{\beta _{1}}{\epsilon _{1} \left( \alpha _{1} ^{2} + \beta _{1} ^{2} \right) + \epsilon _{2} k ^{2} _{\perp} + k _{\perp} \alpha _{1} \left( \epsilon _{1} + \epsilon _{2} \right)} \notag \\ & \times k _{\perp} J _{1} (k _{\perp} \rho) e ^{- k _{\perp} ( z + z ^{\prime} )} dk _{\perp}  . \label{Psi>}
\end{align}
The magnetic field is obtained from these expressions as $\textbf{B} = \nabla \times \textbf{A} = - \left( \partial _{z} \Psi \right) \hat{\textbf{e}} _{\rho} + \rho ^{-1} \partial _{\rho} \left( \rho \Psi \right) \hat{\textbf{e}} _{z}$. In Fig. \ref{EMPlots}c we show the magnetic field $\textbf{B}$ (in units of $q \Sigma _{1} ^{2}$) induced by an electric charge in vacuum at $z ^{\prime} = 1 / \Sigma _{1}$ close to the WSM TaAs as a function of the dimensionless coordinates $\Sigma _{1} \rho$ and $\Sigma _{1} z$. Clearly, the field lines  do not have a simple form. The magnetic field generated by an electric charge close to a TI should serve as the benchmark for understanding the subtlety of our result. In that case, the monopole magnetic field beneath (above) the surface is radially directed with the magnetic monopole above (beneath) the surface as its origin \cite{Qi-Monopole, Karch, MCU-GreenTI}. In the present case, however, the behavior of the field lines is radically different. Above the surface, the magnetic field lines begin at the surface and end at the surface (but not at the same point). The situation beneath the surface also differs from that of the topological insulator. In  Sec. \ref{Conclusions} we discuss two experimental setups which could be used to test this nontrivial magnetic field. 

To understand the physical origin of the induced magnetic field above the surface we rewrite the Maxwell equation (\ref{Ampere}) as $\nabla \times \textbf{B} _{z>0} = \frac{4 \pi}{c} \textbf{J} _{\mbox{\scriptsize bHall}}$, where the bulk Hall current, given by $\textbf{J} _{\mbox{\scriptsize bHall}} = \frac{\alpha c}{2 \pi ^{2}} \textbf{b} \times \textbf{E} _{z<0}$, is induced by the in-plane component of the electric field produced by the charge. Having taken $\textbf{b} = b \hat{\textbf{e}} _{z}$,  the current is circulating around the symmetry axis, i.e. $\textbf{J} _{\mbox{\scriptsize bHall}} = \sigma _{xy}  \left( \hat{\textbf{e}} _{\rho} \cdot \textbf{E} _{z<0} \right) \, \hat{\textbf{e}} _{\varphi}$. In Fig. \ref{HallCurrentPlots} we show a stream density plot of the bulk Hall current $\textbf{J} _{\mbox{\scriptsize bHall}}$ (in units of $q \sigma _{xy} \Sigma _{1} ^{2}$) for $z ^{\prime} = 1 / \Sigma _{1}$ and different values of $\Sigma _{1} z$. We observe that each cross section of the bulk Hall current resembles the surface Hall current induced by an electric charge near to a topological insulator. Naively, this suggests that a 3D Weyl semimetallic phase can be understood as an infinite number of 2+1 Dirac subsystems (one for each value of $z$ in the bulk) supporting a surface Hall current. According to Fig. \ref{EMPlots}c, we do not expect an induced magnetic monopole in the bulk as it is the case of an electric charge in front of a TI. A close inspection to Fig. \ref{EMPlots}c reveals that below the surface of the WSM, centered at the position of the image  charge, the $\mathbf{B}$-field lines wind in an axisymmetric way  as if about a loop of current, similar to those of a ``physical'' magnetic dipole of finite radius. This suggests that we consider a multipole expansion of the magnetic field and determine the dominant contribution. Still, we recall that the source of the magnetic field is not localized, being the bulk Hall current $\textbf{J} _{\mbox{\scriptsize bHall}}$ which is proportional to the  electric field  $\mathbf{E}_{z<0}$  produced by the charge. In this way, the standard multipole expansion for localized sources does not necessarily applies. In order to answer the above  question  we   look for  the large distance behavior of the magnetic potential  $\mathbf{A}=\Psi (\rho ,z)\hat{\mathbf{e}}_{\varphi }$ in the region $z>0$. It is convenient to rewrite Eq. (\ref{Psi>}) in the form $\Psi _{z>0} = \,2q\epsilon _{1}\int_{0}^{\infty }F(k_\perp;\epsilon_1, \epsilon_2)J_{1}(k_\perp \rho )e^{-k_\perp(z+z^{\prime
})}dk_\perp $, where
\begin{align}
& F (k _{\perp} ; \epsilon _{1} , \epsilon _{2}) = % \notag \\ &
 \frac{\beta_1}{\epsilon _{1} \sqrt{k _{\perp} ^{2} \! + \! \Sigma _{1} ^{2}} \! + \! \epsilon _{2} k _{\perp} \! + \! (\epsilon _{1} \! + \! \epsilon _{2}) \alpha_1} .
\end{align}
Due to the exponential factor $e^{-k_\perp(z+z')}$ together with the rapidly oscillating nature of $J_{1}(k_\perp \rho )$ in the far zone, the integral (\ref{Psi>}) is dominated by the behavior of the integrand for small values of $k_\perp$. A series expansion of  $F(k_\perp;\epsilon_1, \epsilon_2)$ in powers of $k_\perp/\Sigma_1$ results in an expansion in powers of $(k_\perp/\Sigma_1)^{1/2}$. The leading terms are:
\begin{align}
\Psi _{z>0}  &\approx 2 q \epsilon _{1} \int _{0} ^{\infty} \left[  \frac{\epsilon _{1}}{\sqrt{2}} \left(\frac{k_\perp}{\Sigma_1}\right)^{1/2} - \frac{\left( \epsilon _{1} + \epsilon _{2} \right) }{2 \epsilon _{1} ^{2}} \left(\frac{ k _{\perp}}{\Sigma _{1}} \right) \right. \notag \\ & \hspace{0.9cm} \left. + O \left(\frac{k_\perp}{\Sigma_1} \right)^{3/2} \right] J_{1}(k_\perp\rho )e^{-k_\perp(z+z^{\prime})}dk_\perp. \label{EQ3}
\end{align}
Just from dimensional arguments, it is clear that the term proportional to  $k_\perp^\sigma$ (with $\sigma >0 $) under the integral yields a contribution of the order $(1/L)^{(\sigma +1)}$, with $L$ being a characteristic length in the integrand. In fact, the required integrals are given in closed form \cite{Gradshteyn}, yielding
\begin{equation}
\mathbf{A}^{(1)} \approx  \frac{\sqrt{2} \, \Gamma(5/2) \,q}{\Sigma_1}\frac{1}{r^{3/2}}P^{-1}_{1/2}(\cos \theta) \hat{\mathbf{e}}_\varphi,
\end{equation}
for the dominant contribution in the far zone, that arises from the term proportional to $k_\perp^{1/2}$ in Eq. (\ref{EQ3}). Here $\theta$ is the angle from the $oz$ axis to the observation 
point, $r \cos \theta = \mathbf{r} \cdot \hat{\mathbf{e}}_z$ and $\mathbf{r} = \rho \hat{\mathbf{e}}_\rho + (z + z ^{\prime}) \hat{\mathbf{e}}_z$ and $r = \sqrt{\rho^2 + (z + z ^{\prime}) ^2}$. 
The associated Legendre function $P^{-1}_{1/2}(x)$ is
\begin{equation}
P^{-1}_{1/2}(x)=\frac{1}{\Gamma(2)}\left(\frac{1-x}{1+x}\right)^{1/2} \!\!\! {}_{2}F _{1} \left(-1/2,\, 3/2 \, ; \,  2 \, ; \,\frac{1-x}{2}\right),
\end{equation} 
where ${}_{2}F _{1} (a, b ; c ; z)$ is the hypergeometric function.
The next term in Eq. (\ref{EQ3}) produces
\begin{equation}
\mathbf{A}^{(2)} \approx
-q\frac{\left( \epsilon _{1}+\epsilon _{2}\right) }{\epsilon
_{1} \Sigma _{1}}\frac{\sin \theta }{r^{2}} \hat{\mathbf{e}}_{\varphi}.
\end{equation}
Comparing with the magnetic potential $\mathbf{A}= \left(m \sin \theta/r^{2}\right)\hat{\mathbf{e}}_{\varphi }$ produced by a magnetic 
dipole  $m \hat{\mathbf{e}} _z$ at the origin,
we identify this contribution as that of a magnetic dipole with
\begin{equation}
m =-q \frac{\left( \epsilon _{1}+\epsilon _{2}\right) }{\epsilon
_{1}\Sigma _{1}},
\end{equation}
located at the image point $-z ^{\prime}$. Thus we confirm the qualitative expectation that  a magnetic dipole is induced, which  is the subleading term of the vector potential.

\section{Interaction energy and force} \label{Force}

\begin{figure}
\includegraphics[scale=0.52]{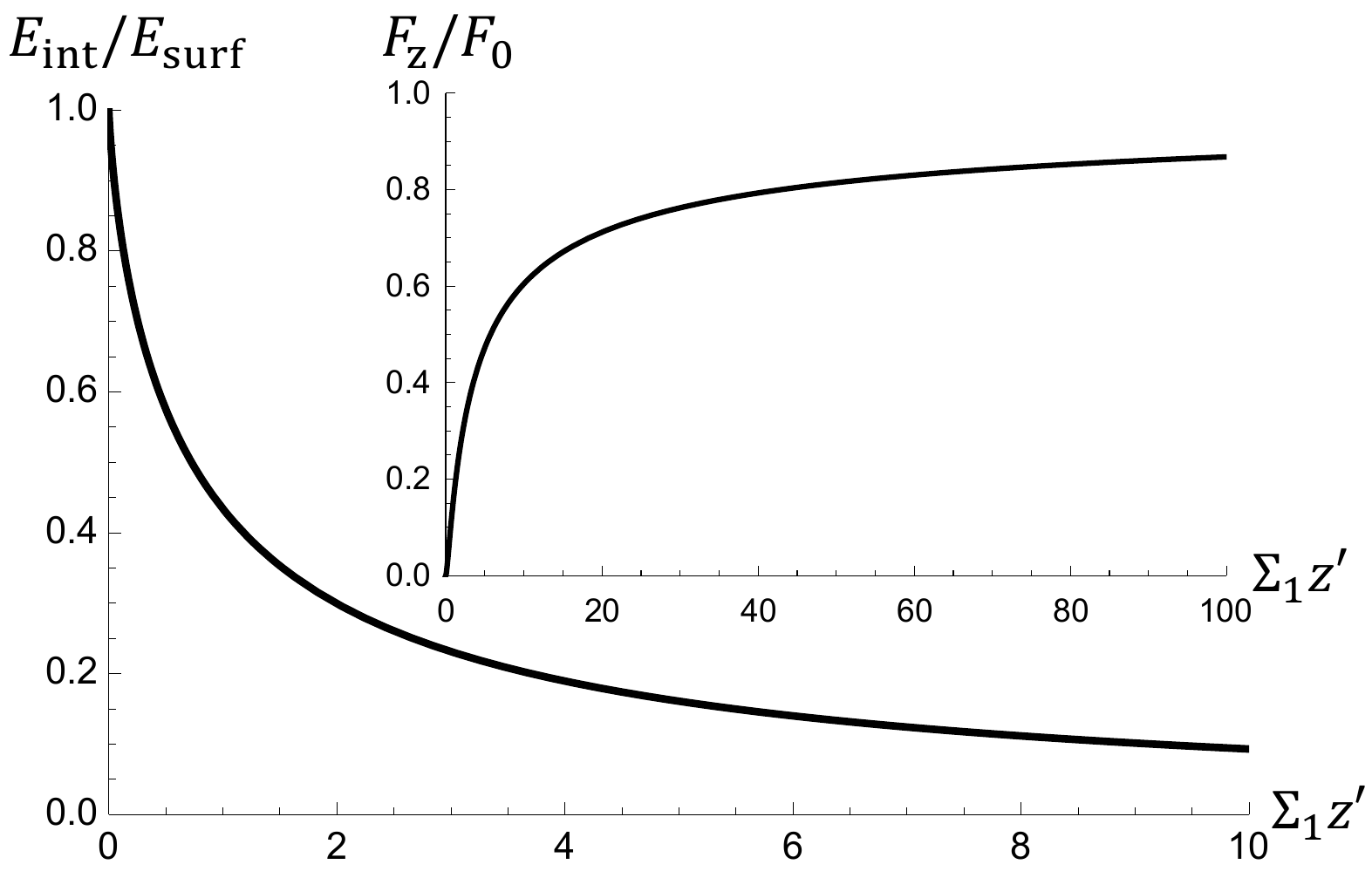}
\caption{\small Interaction energy $E _{\mbox{\scriptsize int}}$ (in units of the surface energy $E _{\mbox{\scriptsize surf}}$) as a function of the dimensionless distance $\Sigma _{1} z ^{\prime}$. The inset shows the force (in units of $F _{0}$) as a function of $\Sigma _{1} z ^{\prime}$.} \label{InteractionEnergy}
\end{figure}

To compute the force between the electric charge and the Weyl semimetal we need the interaction energy between a charge distribution and a WSM as given by \cite{Schwinger}
\begin{align}
E _{\mbox{\scriptsize int}} = \frac{1}{2} \int \left[ \Phi (\textbf{r}) - \Phi _{0} (\textbf{r}) \right] \rho (\textbf{r}) d ^{3} \textbf{r} ,
\end{align}
where $\Phi _{0} (\textbf{r}) = \lim _{\Sigma _{1} \rightarrow 0} \Phi (\textbf{r})$ is the electrostatic potential in the absence of the $\theta$ term. The first contribution represents the total energy of a charge distribution in the presence of the WSM, including mutual interactions. We evaluate this energy for the problem of an electric charge above a WSM. Making use of Eq. (\ref{EscPotDielectric}), the interaction energy becomes
\begin{align}
E _{\mbox{\scriptsize int}} (z ^{\prime}) = & - \frac{q ^{2}}{4 \epsilon _{2} z ^{\prime}} \frac{\epsilon _{1} - \epsilon _{2}}{\epsilon _{1} + \epsilon _{2}} - \frac{q ^{2} \epsilon _{1}}{\epsilon _{1} + \epsilon _{2}} \int _{0} ^{\infty} e ^{-2 k _{\perp} z ^{\prime}} \notag \\ & \times  \frac{\alpha _{1} ^{2} + \beta _{1} ^{2} - k ^{2} _{\perp}}{\epsilon _{1} \left( \alpha _{1} ^{2} + \beta _{1} ^{2} \right) + \epsilon _{2} k ^{2} _{\perp} + k _{\perp} \alpha _{1} \left( \epsilon _{1} + \epsilon _{2} \right)} dk _{\perp} , \label{IntEnergy}
\end{align}
which we interpret as follows. The first term corresponds to the interaction energy between the original charge at $z ^{\prime}$ and the image charge at $- z ^{\prime}$ \cite{Schwinger}. The second term does not admit an immediate interpretation in a similar fashion, however, we are certain that it is a consequence of the nontrivial bulk topology of the material since it vanishes as the bulk Hall conductivity goes to zero. We observe that as the charge approaches the interface ($z ^{\prime} \rightarrow 0$), the nontopological contribution will dominate the  interaction energy (\ref{IntEnergy}) provided $\epsilon _{1} \neq \epsilon _{2}$; and therefore $E _{\mbox{\scriptsize int}} \rightarrow - \infty$, as usual. However, this trivial contribution vanishes for $\epsilon _{1} = \epsilon _{2}$, which is achieved by embedding the charge in a dielectric fluid with the same permittivity to that of the Weyl semimetal. This idea was recently employed in Refs. \cite{MU, MC} to cancel out the trivial electrostatic effects when studying the interaction between an hydrogen-like ion and a planar topological insulator. 
To isolate the topological effects we focus on this case. A distinguishing feature of this interaction energy is that it does not diverge as the charge approaches the interface. Indeed, we can compute the surface interaction energy analytically, with the result (setting $\epsilon _{1} = \epsilon _{2} \equiv \epsilon$)
\begin{align}
E _{\mbox{\scriptsize surf}} \equiv E _{\mbox{\scriptsize int}} (z ^{\prime} = 0) &= - \frac{\alpha q ^{2} b}{8 \epsilon ^{2}}.
\end{align} 
This finite value of the interaction energy at the interface is a signature that the electric field cannot be interpreted in terms of a symmetrically located image charge, as in metals, dielectrics and topological insulators. In Fig. \ref{InteractionEnergy} we show a plot of the ratio between the interaction energy $E _{\mbox{\scriptsize int}}$ and the surface energy $E _{\mbox{\scriptsize surf}}$ as a function of the dimensionless distance $\Sigma _{1} z ^{\prime}$. We observe that the maximum value is precisely at the surface, and it decreases asymptotically to zero as the charge moves away from the surface. The force that the Weyl semimetal exerts upon the charge can be computed as $F _{z} (z ^{\prime}) = - \partial _{z ^{\prime}} E _{\mbox{\scriptsize int}} (z ^{\prime})$. To get an insight of the magnitude of this force, in the inset of Fig. \ref{InteractionEnergy} we plot the force $F _{z}$ (in units of $F _{0} = - \frac{q ^{2}}{ \epsilon (2 z ^{\prime}) ^{2}}$, which is the force that a perfect metallic surface exerts upon the charge) as a function of the dimensionless coordinate $\Sigma _{1} z ^{\prime}$. As we can see, the force between the Weyl semimetal and the charge tends asymptotically to the force between the charge and a perfect metallic surface.

\section{Summary and Discussion} \label{Conclusions}

In summary, we have computed the electromagnetic fields produced by an electric charge near  a topological Weyl semimetal in the equilibrium state, at the neutrality point, and with two nodes in the bulk Brillouin zone, when the charge is located in front of the face without surface states. We found that, outside the WSM, the electric field behaves as that generated by the original electric charge, with deviations close to the interface due to the screening of the field inside the material (see Fig. \ref{EMPlots}b). This behavior is dominated by the dielectric properties of the semimetal, in such a way that the topological contribution is always hidden. The magnetic field is, on the contrary, of topological origin due to the magnetoelectric effect of topological phases. In particular, we showed that the magnetic field exhibits a characteristic behavior above the WSM: the field lines begin at the surface and then end at  the surface (but not at the same point), as depicted in Fig. \ref{EMPlots}c.  In fact, we showed that this peculiar magnetic field for $z \gg z'$ includes a nonleading contribution corresponding to a magnetic dipole moment induced beneath the WSM's surface. This magnetic field is different  from the radially directed one produced by an electric charge near the surface of a TI, interpreted in terms of an image magnetic monopole located beneath  surface. As in the case of the charge in front of the TI, in our case, the interpretation of the dipole magnetic moment is only an artifact. The physical origin of this field are the circulating Hall currents induced in the bulk of the WSM obtained at the end of Sec. \ref{PartSol}. Again, the comparison with the situation of a charge in front of a TI  is useful. As we see from the stream density plot of the bulk Hall current in Fig. \ref{HallCurrentPlots}, for each  $z<0$, the current resembles the surface Hall current induced by a charge near a TI, suggesting that a 3D WSM can be interpreted as an infinite number of 2+1 Dirac subsystems supporting a surface Hall current. The 
distinctive behavior of the magnetic fields here obtained  is an experimentally observable signature of the anomalous Hall effect in the bulk, and thus its detection is in order.

We must recall that our model is based on a simplified description of Weyl semimetals. Nevertheless the  physical realization of materials with generic WSM phases amenable  for experimental measurements is rather subtle. For example,  Weyl semimetals may have more than a single pair \cite{NN1, NN2, NN3} of Weyl nodes and possibly not all aligned with each other. In this case a different approach must be employed to solve the field equations and our results can not be directly applied since axial symmetry no longer holds. We point out that our model and results still apply to systems where the Weyl nodes appear once time-reversal is broken by an external magnetic field. For instance, in the Dirac materials Cd$_{3}$As$_{2}$ \cite{DiracCdAs1, DiracCdAs2, DiracCdAs3} and Na$_{3}$Bi \cite{DiracNaBi1, DiracNaBi2, DiracNaBi3}, each Dirac point is expected to split into two Weyl nodes with a separation in momentum proportional (in magnitude and direction) to the magnetic field. An interesting theoretical proposal of the WSM phase with two nodes is the multilayer structure comprised of topologically trivial and nontrivial insulators proposed in Refs. \cite{Burkov-Balents,Burkov,Burkov2}, which up to our knowledge has not been realized experimentally yet.

Now we discuss two specific fingerprints of the induced magnetic field above our particular WSM  which could, in principle,  be measured.

\emph{Angle-resolved measurement.} The force that the Weyl semimetal exerts upon the charge is $F _{z} = - \partial _{z ^{\prime}} E _{\mbox{\scriptsize int}} (z ^{\prime})$, where the interaction energy is given by Eq. (\ref{IntEnergy}). This force corresponds to $\textbf{F} _{e} = q \textbf{E} _{z>0} ( \textbf{r} ^{\prime})$, where $\textbf{E} _{z>0}$ is the electric field above the WSM evaluated at the position $\textbf{r} ^{\prime} = z ^{\prime} \hat{\textbf{e}} _{z}$ of the original charge, and it attracts the charge toward the surface in the direction perpendicular  to it. However, interesting phenomena appear when we examine a moving external charge. For example, consider a steady electron beam drifting at a distance $z ^{\prime}$ above the surface of the WSM. If the motion of the electrons is slow enough with respect to the Fermi velocity in the solid, 
the induced polarization and magnetization of the material rearranges infinitely fast, in such a way that the solution for the electromagnetic fields we have computed are still valid. In this case, where the charge $q$ is moving with a uniform velocity $\textbf{v}$ above the surface of the WSM, the force acting upon the charge will acquire an additional term of the form $\textbf{F} _{m} = q {\textbf{v}  \over c} \times \textbf{B} _{z>0} (\textbf{r} ^{\prime})$ due to the induced magnetic field. For an electron beam moving along the $x$-direction (with velocity $\textbf{v} = v _{x} \hat{\textbf{e}} _{x}$) we find
\begin{align}
\textbf{F} _{m} = - \hat{\textbf{e}} _{y} 
\int _{0} ^{\infty} \!\!\!\! \frac{2 q ^{2} \epsilon _{1} (v _{x}/c) \; k ^{2} _{\perp} \beta _{1} e ^{- 2 k _{\perp} z ^{\prime}} dk _{\perp}  }{\epsilon _{1} \left( \alpha _{1} ^{2} + \beta _{1} ^{2} \right) + \epsilon _{2} k ^{2} _{\perp} + k _{\perp} \alpha _{1} \left( \epsilon _{1} + \epsilon _{2} \right)}. \label{Force2}
\end{align}
Remarkably, this anomalous force is orthogonal to the electrons' motion as well as to the electric contribution $\textbf{F} _{e}$. As a result, these effects can be distinguished from each other. Experimentally, the required probe can be provided by the steady electron beam emitted from a low-energy electron gun (low-energy electron diffraction). While drifting above the WSM, the anomalous force (\ref{Force2}) will deflect the trajectory of the electron beam. To estimate the size of this deflection we consider the proposal in Ref. \cite{ZangNagaosa} of a similar experimental setup involving TIs. In that case the authors take $v _{x} \sim 10 ^{7}$ cm/s, $z ^{\prime} \sim 1 \, \mu$m and $L \sim 1$ cm for the sample's size (which in their case coincides with the $ox$-displacement).  We assume these parameters are also feasible when the sample is a WSM.  An estimate of the transverse displacement $\Delta$ produced by the anomalous force $F_{m}$ is
\begin{align}
\Delta  & \approx  \frac{4 \alpha ^{2}}{\pi ^{2} \epsilon _{1}} \, \frac{q ^{2} L ^{2} b ^{2}}{m _{e} v _{x}c } \, f(\epsilon _{1} , \epsilon _{2}; \Sigma z ^{\prime}), \label{AnomalousDrift}
\end{align}
where $f(\epsilon _{1} , \epsilon _{2}; \Sigma z ^{\prime})$ is obtained from  Eq. (\ref{Force2}). Taking $\epsilon_1 \sim 6$ and $b \sim 10^9\,  \mathrm{m}^{-1}$  as for the genuine Weyl semimetal TaAs \cite{TaAs-Huang, TaAs-Lv, TaAs-Xu, TaAs-Yang, TaAs-Xu2}, we find $\Delta \approx 3,2 \, \mu$m. This deflection is 
of the same order of magnitude as that reported in Ref. \cite{ZangNagaosa}, and then it can be traced by angle-resolved measurements.

If this experiment were carried out with a Dirac semimetal by applying an external magnetic field, instead of a genuine WSM, the induced magnetic field will be overwhelmed by the external one, and so would its contribution to the Lorentz force on a moving charge. In fact, we can estimate by how much is the Lorentz force of the external field larger than that of the anomalous force. Taking $\textbf{B} _{\scriptsize \mbox{ext}} = B _{0} \hat{\textbf{e}} _{z}$, the Lorentz force $\textbf{F} _{0} = q \frac{\textbf{v}}{c} \times \textbf{B} _{\scriptsize \mbox{ext}}$ will deflect the trajectory of each electron by an amount $\Delta _{0} \approx \frac{qL ^{2}}{2m v _{x} c} B _{0}$. Thus, the total transverse drift will be $\Delta _{0} + \Delta $, where $\Delta$ is the drift in Eq. (\ref{AnomalousDrift}) produced by the anomalous force. To compare the magnitudes of these contributions we focus on $\Delta / \Delta _{0}$, which is
\begin{align}
\frac{\Delta}{\Delta _{0}} = \frac{8 \alpha ^{2}}{ \pi ^{2} \epsilon _{1}} \frac{qb ^{2}}{B _{0}} f(\epsilon_1, \epsilon_2; \Sigma _{1} z') .
\end{align}
A numerical estimation is obtained by considering the Dirac semimetal Cd$_{3}$As$_{2}$ in the presence of a magnetic field of $1$T. According to Ref. \cite{DiracCdAs3}, this induces a separation of the Weyl nodes of $b = 5 \times 10 ^{8} \mbox{m} ^{-1}$ and $\epsilon _{1} = 12$. Therefore for an electron at a distance $z ^{\prime} = 1 \mu$m above the material we obtain $\Delta / \Delta _{0} \approx 10 ^{-7}$, thus implying that the external field overwhelms the topological contribution by this factor. We then conclude that an angle-resolved measurement is appropriate for experimental realization only if it were possible to consider a genuine WSM, for which no external magnetic field is needed.

\begin{figure}
\includegraphics[scale=0.46]{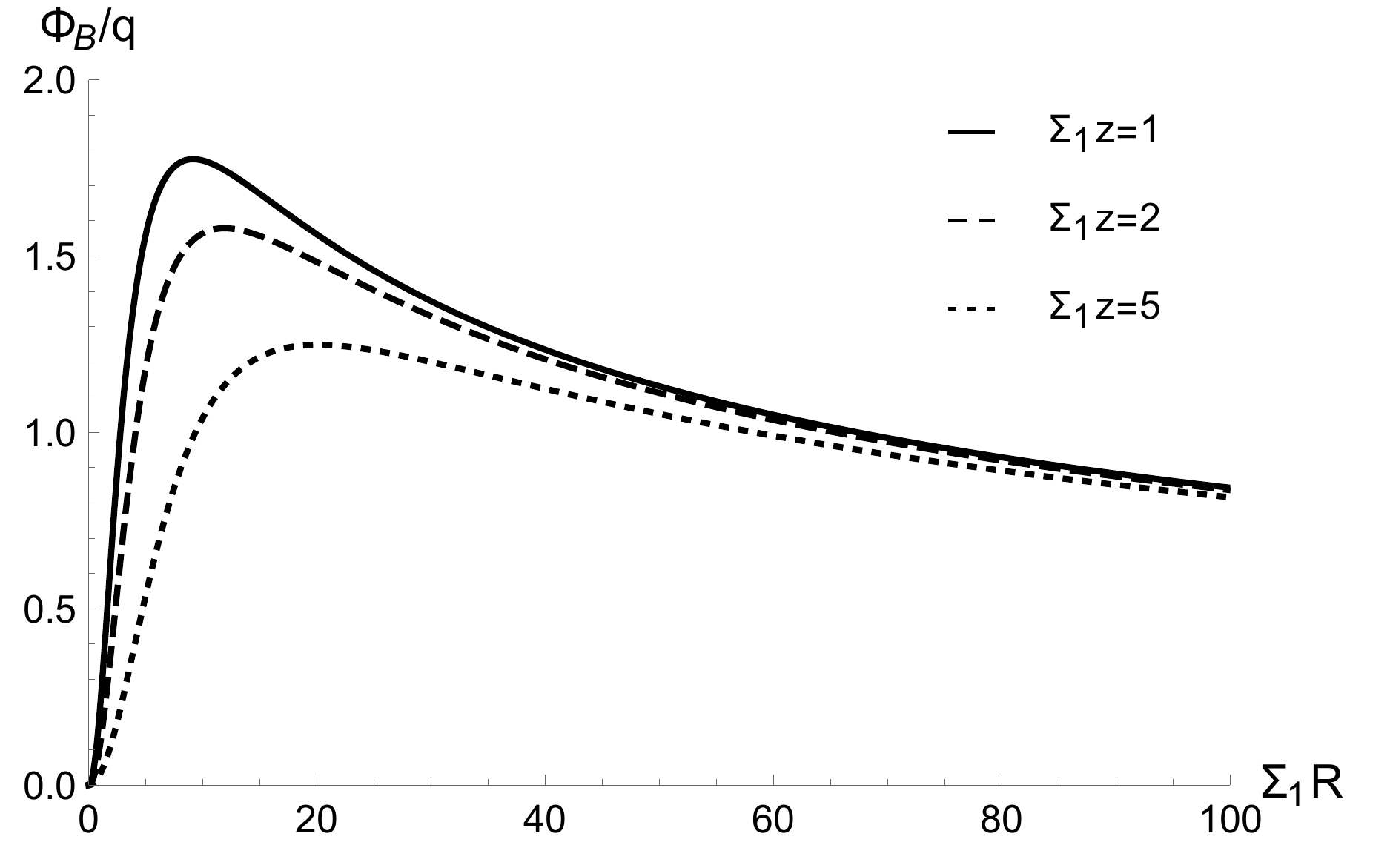}
\caption{\small Magnetic flux generated by an electric charge $q$ at a distance $z ^{\prime} = 1 / \Sigma _{1}$ above the Weyl semimetal TaAs as a function of $\Sigma _{1} R$ for different values of $\Sigma _{1} z$.}
\label{FluxPlot}
\end{figure}

\emph{Scanning SQUID magnetometry.} Another possible technique for measuring the induced magnetic field could be scanning SQUID (Superconducting Quantum Interference Device) magnetometry.  Roughly speaking, a SQUID is a very sensitive magnetometer used to measure extremely subtle magnetic fields (as low as $5 \times 10 ^{-18}$T \cite{SQUID1, SQUID2, SQUID3}), based on superconducting loops containing Josephson junctions. Technically, we have to compute the magnetic flux through a loop (of radius $R$ and parallel to the surface) placed at a distance $z$ above the Weyl semimetal, i.e. $\Phi _{\textbf{B}} = \int _{S} \textbf{B} \cdot d \textbf{S}$, where $S$ is the surface of the loop. The magnetic flux from a topological insulator through a Josephson junction, $\Phi _{\textbf{B}} ^{\mbox{\scriptsize{TI}}}$, serves as the benchmark for comparing our result. In that case, the magnetic flux grows from 0 (at $R=0$) to the constant value $2 \pi g$  (as $R \rightarrow \infty$), where $g = \frac{2q \alpha}{2 (\epsilon _{1} + \epsilon _{2}) + \alpha ^{2}}$ is the magnetic monopole strength \cite{MCU-GreenTI}. This is so because the magnetic field is radially directed away from the image magnetic monopole beneath the surface and therefore the loop will always enclose field lines. This interesting tendency of the magnetic flux to a constant value can be thought as a distinctive feature of the induced magnetic field in topological insulators. The case of a Weyl semimetal is quite different, as we discuss below.
\begin{figure}
\includegraphics[scale=0.45]{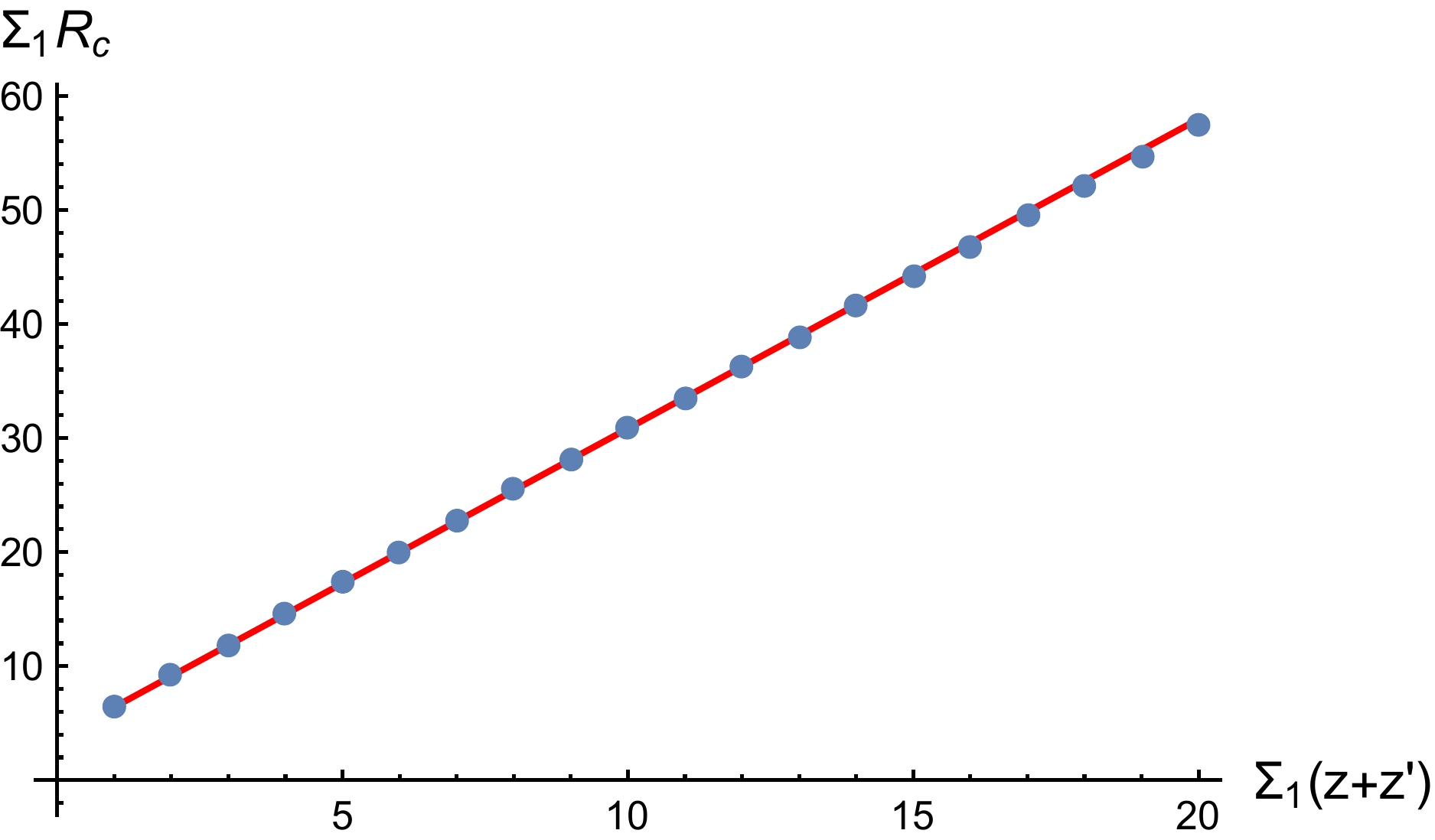}
\caption{\small Critical radius $\Sigma _{1} R _{c}$ of the loop corresponding to the maximal magnetic field flux as a function of $\Sigma _{1} (z+z ^{\prime})$ for the Weyl semimetal TaAs. The dots correspond to the numerical calculation and the continuous line is a curve-fitting.}
\label{CriticalR}
\end{figure}

In our case, a simple calculation produces $\Phi _{\textbf{B}} (R , z) = 2 \pi R \Psi _{z > 0} (R , z)$, where the function $\Psi _{z > 0}$ is given by Eq. (\ref{Psi>}). In Fig. \ref{FluxPlot} we show a plot of the magnetic flux $\Phi _{\textbf{B}} (R , z)$ (in units of $q$) due to a pointlike charge $q$ located at a distance $z ^{\prime} = 1/ \Sigma _{1}$ above the surface of the Weyl semimetal TaAs (for which $\epsilon _{1} \sim 6$ and $b \sim 10 ^{9} \mbox{m} ^{-1}$) as a function of the dimensionless radius $\Sigma _{1} R$ and for different values of $\Sigma _{1} z$. Of course, $\Phi _{\textbf{B}} = 0$ at $R = 0$. Furthermore, in the limit $R \to \infty$, the function $\Psi _{z>0}$ (\ref{Psi>}) is a highly oscillatory integral and therefore $\Phi _{\textbf{B}} \to 0$. This behavior implies the existence of a maximum flux at a critical radius $R _{c}$, as shown in Fig. \ref{FluxPlot}. The fact that the magnetic flux tends to zero as the radius goes to infinity can be easily understood from the fact that the magnetic field lines, which start at the WSM surface, go back again to the surface, as discussed before. The existence of a maximum flux at $R _{c}$, as well as the asymptotic vanishing of the flux, are distinguishing features of the induced magnetic field due to a genuine WSM. One can further determine the critical radius $R _{c}$ corresponding to the maximal magnetic field flux in the usual manner (i.e. by solving $\partial  _{R} \Phi _{\textbf{B}} \vert _{R = R _{c}} = 0$ for $R _{c}$). In Fig. \ref{CriticalR} we show a plot of $\Sigma _{1} R _{c}$ as a function of $\Sigma _{1} (z+z ^{\prime})$ for the case in which the WSM is the TaAs. While the dots represents the numerical solution of $\partial  _{R} \Phi _{\textbf{B}} \vert _{R _{c}} = 0$, the continuous line is a curve-fitting. Unexpectedly we find the equation of a straight line, $\Sigma _{1} R _{c} = m \, \Sigma _{1} (z+z ^{\prime}) + b$, where the values of the slope $m$ and the intercept $b$ on the $\Sigma _{1} R _{c}$-axis depend on the permittivity $\epsilon _{2}$ of the WSM. When the WSM is the TaAs, we find $m = 2.71$ and $b = 3.69$. For a numerical estimate of the magnetic flux we consider a charge $q = n _{e} \vert e \vert$ placed at a distance $z ^{\prime} = 1 \mu$m above the surface of the Weyl semimetal TaAs and a SQUID of radius $R = 10 \mu$m located at $z = 10 \mu$m. We find $\Phi _{\textbf{B}} \approx 7 n _{e} \times 10 ^{-14} \mbox{T} \cdot \mbox{cm} ^{2}$, which is measurable with present day attainable sensitivities of SQUIDs \cite{SQUID1, SQUID2, SQUID3}. One of the key challenges for the experimental detection of this flux profile would be to find a way to fix and localize the charge above the surface.

As expected, if this experiment were carried out with a Dirac semimetal instead of a genuine WSM, the required external magnetic field will overwhelm the topological contribution to the total magnetic flux. Nevertheless, in this case it is still possible to disentangle these effects by using the fact that the contribution to the flux produced by the external magnetic field, say $\Phi _{\textbf{B}} ^{\scriptsize \mbox{ext}}$, is constant in space and time. Contrary to this, the contribution from the induced field $\Phi _{\textbf{B}}$ is also constant in time, however, it is not isotropic. A sensitive magnetometer as the SQUID will be capable to measure small variations of the flux which amounts to eventually measuring the induced electromotive force $\mathcal{E}$ in the loop. Therefore, this allows for isolating the topological contribution, for example, by producing a controlled displacement of the SQUID along the $z$-axis at speed $v _{z}$, namely: $\mathcal{E} = - \frac{d}{dt} \left( \Phi _{\textbf{B}} ^{\scriptsize \mbox{ext}} + \Phi _{\textbf{B}} \right) = - v _{z} \frac{d \Phi _{\textbf{B}}}{dz}$, where the $z$-dependence is read-off from Eq. (\ref{Psi>}).

\acknowledgments

We thank Alberto Cortijo for useful comments and suggestions, and also to the anonymous referees for their recommendations. A. M. was supported by the CONACyT postdoctoral Grant No. 234774. L.F.U. has been supported in part by the project CONACyT (M\'{e}xico) \# 237503.  M.C. has been partially supported by  UNAB DGID under grant \# DI-33-17/RG and wishes to thank A. Mart\'in-Ruiz and L. F. Urrutia at Instituto de Ciencias Nucleares, UNAM for the kind hospitality of throughout the preparation of the manuscript.

\appendix

\section{Detailed solution} \label{DetSol}

In this section we present the detailed solution of the equations of motion Eqs. (\ref{MaxRed1}) and (\ref{MaxRed2}) in the general case where two bulk Hall materials are in contact, as described in the main text. First we derive the corresponding boundary conditions at the interface $z = 0$ and at the singular point $z = z ^{\prime}$; and next we use standard electromagnetic techniques to obtain the general solution in the whole space.

\emph{Boundary Conditions}. The boundary conditions for $\phi$ and $\psi$ can be determined as usual. Assuming that the reduced functions are bounded when $z$ is in the infinitesimal neighborhood of $z = 0$, integration of Eqs. (\ref{MaxRed1}) and (\ref{MaxRed2}) over the interval between $- \varepsilon$ and $+ \varepsilon$ with $\varepsilon \rightarrow 0 ^{+}$, yields the continuity of $\epsilon \partial _{z} \phi$ and $\partial _{z} \psi$ there. Then, the continuity of $\phi$ and $\psi$ at $z = 0$ follows. In a similar fashion, one can show that the singularity in Eq. (\ref{MaxRed1}) requires, at $z = z ^{\prime}$, that $\phi$ be continuous, while $\partial _{z} \phi$ be discontinuous, i.e. $\partial _{z} \phi \vert _{z = z ^{\prime} - 0} ^{z = z ^{\prime} + 0}  = - q / \epsilon (z ^{\prime})$. Analogously, integrating twice Eq. (\ref{MaxRed2}) yields the continuity of $\psi$ and $\partial _{z} \psi$ at the singular point. 

\emph{General solution}. The solutions to equations (\ref{MaxRed1}) and (\ref{MaxRed2}) subject to the above  boundary conditions can be expressed in terms of the solutions, $e ^{k _{jz} z}$, $e ^{k _{jz} ^{\ast} z}$, $e ^{- k _{jz} z}$ and $e ^{- k _{jz} ^{\ast} z}$, of the corresponding homogeneous equations. Here, $j$ labels the two media and $k _{jz} = \alpha _{j} + i \beta _{j}$ is the complex wave number, with $\alpha _{j}$ and $\beta _{j}$ given by Eq. (\ref{kappa}).

To compute the reduced functions $\phi$ and $\psi$  we first partition the $z$-axis in the three regions: (I) $z<0$, (II) $0 < z < z ^{\prime}$ and (III) $z ^{\prime} < z$. Next, we write an appropriate linear combination of the solutions to the homogeneous equation for each region, and finally we apply the corresponding boundary conditions. On the one hand, for the reduced scalar potential $\phi$, the forms of the solutions in the three regions are as follows:
\begin{align}
\phi _{I} &= a _{1} e ^{k _{1z} z} + a _{2} e ^{k _{1z} ^{\ast} z} \label{phiI} , \\ \phi _{II} &= b _{1} e ^{ k _{2z} z} + b _{2} e ^{ k _{2z} ^{\ast} z} + c _{1} e ^{ - k _{2z} z} + c _{2} e ^{ - k _{2z} ^{\ast} z} , \label{phiII} \\ \phi _{III} &= d _{1} e ^{ - k _{2z} z} + d _{2} e ^{ - k _{2z} ^{\ast} z} , \label{phiIII}
\end{align}
where the signs in the exponentials (\ref{phiI}) and (\ref{phiIII}) are required by the boundary condition that $\phi$ goes to zero for $\vert z \vert \rightarrow \infty$. On the other hand, we observe that Eq. (\ref{MaxRed2}) dictates the relation between $\psi$ and $\phi$; namely, $\psi \sim i \epsilon _{j} \phi$ for $\phi \sim e ^{\pm k _{jz} z}$ and $\psi \sim - i \epsilon _{j} \phi$ for $\phi \sim e ^{\pm k _{jz} ^{\ast} z}$. Using this result we find that Eq. (\ref{phiI}) implies that, for the region I, $\psi _{I} = i \epsilon _{1} \left( a _{1} e ^{ k _{1z} z} - a _{2} e ^{ k _{1z} ^{\ast} z} \right)$. In a similar fashion we obtain the corresponding expressions for $\psi _{II}$ and $\psi _{III}$.

Imposing the boundary conditions and solving the resulting system of equations we find for the coefficients
\begin{widetext}
\begin{align}
a _{1} &= a _{2} ^{\ast} = \frac{q}{2 \epsilon _{1}} \frac{k _{2z} \left( \epsilon _{1} - \epsilon _{2} \right) e ^{- k _{2z} ^{\ast} z ^{\prime}} + \left[ k _{2z} ^{\ast} \left( \epsilon _{1} + \epsilon _{2} \right) + 2 k _{1z} ^{\ast} \epsilon _{1} \right] e ^{- k _{2z} z ^{\prime}}}{2 \left( \epsilon _{1} k _{1z} k _{1z} ^{\ast} + \epsilon _{2} k _{2z} k _{2z} ^{\ast} \right) + \left( \epsilon _{1} + \epsilon _{2} \right) \left( k _{1z} ^{\ast} k _{2z} + k _{1z} k _{2z} ^{\ast} \right) } , \\ b _{1} &= b _{2} ^{\ast} = \frac{q}{4 \epsilon _{2} k _{2z}} e ^{- k _{2z} z ^{\prime}} , \\ c _{1} &= c _{2} ^{\ast} = \frac{q}{4 \epsilon _{2} k _{2z}} \frac{\left[ 2 \left( \epsilon _{2} k _{2z} k _{2z} ^{\ast} - \epsilon _{1} k _{1z} k _{1z} ^{\ast} \right) + \left( \epsilon _{1} + \epsilon _{2} \right) \left( k _{2z} k _{1z} ^{\ast} - k _{2z} ^{\ast} k _{1z} \right) \right] e ^{- k _{2z} z ^{\prime}} - 2 k _{1z} k _{2z} \left( \epsilon _{1} - \epsilon _{2} \right) e ^{- k _{2z} ^{\ast} z ^{\prime}} }{2 \left( \epsilon _{1} k _{1z} k _{1z} ^{\ast} + \epsilon _{2} k _{2z} k _{2z} ^{\ast} \right) + \left( \epsilon _{1} + \epsilon _{2} \right) \left( k _{1z} ^{\ast} k _{2z} + k _{1z} k _{2z} ^{\ast} \right)} , \\ d _{1} &= d _{2} ^{\ast} = \frac{q}{4 \epsilon _{2} k _{2z}} \left\lbrace e ^{ k _{2z} z ^{\prime}} \!\!  + \!  \frac{\left[ 2 \left( \epsilon _{2} k _{2z} k _{2z} ^{\ast} - \epsilon _{1} k _{1z} k _{1z} ^{\ast} \right) + \left( \epsilon _{1} + \epsilon _{2} \right) \left( k _{2z} k _{1z} ^{\ast} - k _{2z} ^{\ast} k _{1z} \right) \right] e ^{- k _{2z} z ^{\prime}} \! \!  - 2 k _{1z} k _{2z} \left( \epsilon _{1} - \epsilon _{2} \right) e ^{- k _{2z} ^{\ast} z ^{\prime}}}{2 \left( \epsilon _{1} k _{1z} k _{1z} ^{\ast} + \epsilon _{2} k _{2z} k _{2z} ^{\ast} \right) + \left( \epsilon _{1} + \epsilon _{2} \right) \left( k _{1z} ^{\ast} k _{2z} + k _{1z} k _{2z} ^{\ast} \right)}  \right\rbrace .
\end{align}
Using these results, we can write the reduced functions beneath the surface as $\phi _{I} = 2 \mbox{Re} \left( a _{1} e ^{k _{1z} z} \right)$ and $\psi _{I} = - 2 \epsilon _{1} \mbox{Im} \left( a _{1} e ^{k _{1z} z } \right)$, whose explicit forms are
\begin{align}
\phi _{I} = & \; \frac{q}{\epsilon _{1} Q} e ^{\alpha _{1} z - \alpha _{2} z ^{\prime}} \Big[ \left( \epsilon _{1} \alpha _{1} + \epsilon _{2} \alpha _{2} \right) \cos \left( \beta _{1} z - \beta _{2} z ^{\prime} \right) + \left( \epsilon _{1} \beta _{1} + \epsilon _{2} \beta _{2} \right) \sin \left( \beta _{1} z - \beta _{2} z ^{\prime} \right) + \left( \epsilon _{1} - \epsilon _{2} \right) \notag \\ & \times \cos \left( \beta _{1} z \right) \left[ \alpha _{2} \cos \left( \beta _{2} z ^{\prime} \right) - \beta _{2} \sin \left( \beta _{2} z ^{\prime} \right) \right] \Big] , \\ \psi _{I} = & \; \frac{q}{Q} e ^{\alpha _{1} z - \alpha _{2} z ^{\prime}} \Big[ \left( \epsilon _{1} \beta _{1} + \epsilon _{2} \beta _{2} \right) \cos \left( \beta _{1} z - \beta _{2} z ^{\prime} \right) - \left( \epsilon _{1} \alpha _{1} + \epsilon _{2} \alpha _{2} \right) \sin \left( \beta _{1} z - \beta _{2} z ^{\prime} \right) + \left( \epsilon _{1} - \epsilon _{2} \right) \notag \\ & \times \sin \left( \beta _{1} z \right) \left[ \beta _{2} \sin \left( \beta _{2} z ^{\prime} \right) - \alpha _{2} \cos \left( \beta _{2} z ^{\prime} \right) \right] \Big] ,
\end{align} 
which are the ones we present in the main text in Eqs. (\ref{RedEscz<0}) and (\ref{RedVecz<0}). Now we follow similar steps to derive the reduced functions $\phi _{II} = 2 \mbox{Re} \left( b _{1} e ^{k _{2z} z} + c _{1} e ^{- k _{2z} z} \right)$ and $\psi _{II} = - 2 \epsilon _{2} \mbox{Im} \left( b _{1} e ^{k _{2z} z} + c _{1} e ^{- k _{2z} z} \right)$. The results are
\begin{align}
\phi _{II} = & \; \frac{q}{2 \epsilon _{2} r _{2} ^{2}} e ^{\alpha _{2} (z - z ^{\prime})} \Big\{ \alpha _{2} \cos \left[ \beta _{2} \left( z - z ^{\prime} \right) \right] + \beta _{2} \sin \left[ \beta _{2} \left( z - z ^{\prime} \right) \right] \Big\} - \frac{q (\epsilon _{1} - \epsilon _{2})}{2 \epsilon _{2} Q} e ^{- \alpha _{2} (z + z ^{\prime})} \Big\{ \alpha _{1} \cos \left[ \beta _{2} \left( z - z ^{\prime} \right) \right] \notag \\ & + \beta _{1} \sin \left[ \beta _{2} \left( z - z ^{\prime} \right) \right] \Big\} + \frac{q}{2 \epsilon _{2} r _{2} ^{2} Q} e ^{- \alpha _{2} (z + z ^{\prime})} \Big\{ \Gamma \cos \left[ \beta _{2} \left( z + z ^{\prime} \right) \right] - \Delta \sin \left[ \beta _{2} \left( z + z ^{\prime} \right) \right] \Big\} , \\ \psi _{II} = & \; \frac{q}{2 r _{2} ^{2}} e ^{\alpha _{2} (z - z ^{\prime})} \Big\{ \beta _{2} \cos \left[ \beta _{2} \left( z - z ^{\prime} \right) \right] - \alpha _{2} \sin \left[ \beta _{2} \left( z - z ^{\prime} \right) \right] \Big\} + \frac{q (\epsilon _{1} - \epsilon _{2})}{2 Q} e ^{- \alpha _{2} (z + z ^{\prime})} \Big\{ \beta _{1} \cos \left[ \beta _{2} \left( z - z ^{\prime} \right) \right] \notag \\ & - \alpha _{1} \sin \left[ \beta _{2} \left( z - z ^{\prime} \right) \right] \Big\} + \frac{q}{2 r _{2} ^{2} Q} e ^{- \alpha _{2} (z + z ^{\prime})} \Big\{ \Delta \cos \left[ \beta _{2} \left( z + z ^{\prime} \right) \right] + \Gamma \sin \left[ \beta _{2} \left( z + z ^{\prime} \right) \right] \Big\} ,
\end{align}
where $\Gamma$, $\Delta$ and $Q$ are defined in Eq. (\ref{Defs}). Finally, for $\phi _{III} = 2 \mbox{Re} \left( d _{1} e ^{- k _{2z}  z} \right)$ and $\psi _{III} = - 2 \epsilon _{2} \mbox{Im} \left( d _{1} e ^{- k _{2z}  z} \right)$, we obtain
\begin{align}
\phi _{III} = & \; \frac{q}{2 \epsilon _{2} r _{2} ^{2}} e ^{\alpha _{2} (z ^{\prime} - z)} \Big\{ \alpha _{2} \cos \left[ \beta _{2} \left( z ^{\prime} - z \right) \right] + \beta _{2} \sin \left[ \beta _{2} \left( z ^{\prime} - z \right) \right] \Big\} - \frac{q (\epsilon _{1} - \epsilon _{2})}{2 \epsilon _{2} Q} e ^{- \alpha _{2} (z + z ^{\prime})} \Big\{ \alpha _{1} \cos \left[ \beta _{2} \left( z - z ^{\prime} \right) \right] \notag \\ & + \beta _{1} \sin \left[ \beta _{2} \left( z - z ^{\prime} \right) \right] \Big\} + \frac{q}{2 \epsilon _{2} r _{2} ^{2} Q} e ^{- \alpha _{2} (z + z ^{\prime})} \Big\{ \Gamma \cos \left[ \beta _{2} \left( z + z ^{\prime} \right) \right] - \Delta \sin \left[ \beta _{2} \left( z + z ^{\prime} \right) \right] \Big\} , \\ \psi _{III} = & \; \frac{q}{2 r _{2} ^{2}} e ^{\alpha _{2} (z ^{\prime} - z)} \Big\{ \beta _{2} \cos \left[ \beta _{2} \left( z - z ^{\prime} \right) \right] + \alpha _{2} \sin \left[ \beta _{2} \left( z - z ^{\prime} \right) \right] \Big\} + \frac{q (\epsilon _{1} - \epsilon _{2})}{2 Q} e ^{- \alpha _{2} (z + z ^{\prime})} \Big\{ \beta _{1} \cos \left[ \beta _{2} \left( z - z ^{\prime} \right) \right] \notag \\ & - \alpha _{1} \sin \left[ \beta _{2} \left( z - z ^{\prime} \right) \right] \Big\} + \frac{q}{2 r _{2} ^{2} Q} e ^{- \alpha _{2} (z + z ^{\prime})} \Big\{ \Delta \cos \left[ \beta _{2} \left( z + z ^{\prime} \right) \right] + \Gamma \sin \left[ \beta _{2} \left( z + z ^{\prime} \right) \right] \Big\} .
\end{align}
We observe that the reduced functions above the WSM surface can be written in a unified fashion. For example, $\phi _{II}$ and $\phi _{III}$ are both contained in Eq. (\ref{RedEscz>0}); and $\psi _{II}$ and $\psi _{III}$ are both contained in Eq. (\ref{RedVecz>0}).
\end{widetext}

\end{document}